\documentclass[prd,superscriptaddress,amsfonts,amssymb,amsmath,showpacs,twocolumn,nofootinbib]{revtex4-2}
\usepackage{bm}
\usepackage{amsfonts}
\usepackage{latexsym}
\usepackage{graphicx}
\usepackage{amsmath}
\usepackage{palatino}
\usepackage{xcolor} 
\usepackage{mathpazo}
\usepackage{xcolor}
\usepackage{rotating}
\usepackage{adjustbox}
\usepackage{tensor}
\usepackage{textcomp}
\usepackage{float}
\usepackage{booktabs}
\usepackage{dcolumn}
\usepackage[titletoc]{appendix}
\usepackage{booktabs}
\usepackage{multirow}
\usepackage{hyperref}
\hypersetup{colorlinks,citecolor=blue}
\usepackage{amsmath}
\usepackage{xcolor}
\usepackage{orcidlink}
\usepackage{epsfig}
\usepackage{caption}
\usepackage{subcaption}
\usepackage{commath}
\hypersetup{colorlinks,citecolor=blue}
\hypersetup{colorlinks=true,linkcolor=magenta,filecolor=magenta,    urlcolor=blue}

\def\be{\begin{equation}}
\def\ee{\end{equation}}
\def\bea{\begin{eqnarray}}
\def\eea{\end{eqnarray}}

\definecolor{owngreen}{rgb}{0.0, 0.5, 0.0}

\begin{document}

\title{Probing departures from $\Lambda$CDM  by late-time datasets}

\author{Himanshu Chaudhary}
\email{himanshu.chaudhary@ubbcluj.ro,\\
himanshuch1729@gmail.com}
\affiliation{Department of Physics, Babeș-Bolyai University, Kogălniceanu Street, Cluj-Napoca, 400084, Romania}
\affiliation{Research Center of Astrophysics and Cosmology, Khazar University, Baku, AZ1096, 41 Mehseti Street, Azerbaijan}
\author{Vipin Kumar Sharma}
\email{vipinkumar.sharma@iiap.res.in}
\affiliation{Indian Institute of Astrophysics, Koramangala II Block, Bangalore 560034, India}
\author{Salvatore Capozziello}
\email{capozziello@na.infn.it}
\affiliation{Dipartimento di Fisica ``E. Pancini", Universit\`a di Napoli ``Federico II", Complesso Universitario di Monte Sant’ Angelo, Edificio G, Via Cinthia, I-80126, Napoli, Italy,}
\affiliation{Istituto Nazionale di Fisica Nucleare (INFN), sez. di Napoli, Via Cinthia 9, I-80126 Napoli, Italy,}
\affiliation{Scuola Superiore Meridionale, Largo S. Marcellino, I-80138, Napoli, Italy.}
\author{G. Mustafa}
\email{gmustafa3828@gmail.com}
\affiliation{Department of Physics,
Zhejiang Normal University, Jinhua 321004, People’s Republic of China}

\begin{abstract}
Observational data play a pivotal role in identifying cosmological models that are both theoretically consistent and empirically viable. In this work, we investigate the level of preference for dynamical dark energy over a cosmological constant using current late-time observational datasets, including Cosmic Chronometers , Baryon Acoustic Oscillations from DESI DR2, and different Type Ia supernova catalogs (Pantheon$^+$, DES-Dovekie, Union3). We analyze various dynamical dark energy models, including $\omega$CDM, o$\omega$CDM, $\omega_0\omega_a$CDM, Logarithmic, Exponential, JBP, BA, and GEDE. In most cases, the o$\Lambda$CDM and o$\omega$CDM models favor an open Universe. For the o$\omega$CDM, the inclusion of DES-Dovekie or Union3 data together with CC and DESI DR2 favors a nearly flat geometry. Using the CC + DESI DR2 dataset, the preference for dynamical dark energy lies between the $1$-$2\sigma$ level. When different supernova catalogs (DES-Dovekie or Union3) are included, the deviation from $\Lambda$CDM in the $\omega$CDM, $\omega_0\omega_a$CDM, Logarithmic, JBP, BA, and GEDE models increases to the $2$-$2.74\sigma$ level, while the Pantheon$^{+}$ sample yields deviations below the $2\sigma$ level. We find consistent evidence for $\omega_0 > -1$ and $\omega_a < 0$ across all dark energy models, indicating a preference for dynamical dark energy characterized by a Quintom-B type scenario. The $\Lambda$CDM paradigm has long served as the standard framework of modern cosmology; however recent DESI DR2 results have exposed emerging tensions with the cosmological constant $\Lambda$, hinting at possible new physics in the dark energy sector. Even so, the currently available data are still not strong enough to definitively rule out the $\Lambda$CDM model.
\end{abstract}

\maketitle

\section{Introduction}\label{sec_1} 
The late-time accelerated expansion of the Universe remains one of the most profound mysteries in modern cosmology \citep{riess1998observational,SupernovaCosmologyProject:1998vns}. Within the standard Lambda Cold Dark Matter ($\Lambda$CDM) framework, this phenomenon is attributed to a positive cosmological constant $\Lambda$ with negative pressure, introduced in Einstein’s field equations of General Relativity \citep{Einstein:1917ce}. Although $\Lambda$CDM provides an excellent fit to observational data, the cosmological constant itself is a phenomenological parameter lacking a fundamental theoretical explanation for its observed value \citep{Weinberg:1987dv,YaBZeldovich_1968,Weinberg:1988cp}. Furthermore, the magnitude of $\Lambda$ required by current observations implies a cosmic coincidence, marking our epoch as a particularly special time in the evolution of the Universe \citep{Weinberg:1988cp,carroll2001cosmological,Padmanabhan:2002ji}. In response to these conceptual challenges, numerous theoretical models have been proposed to explain cosmic acceleration through dynamical mechanisms \citep{carroll2001cosmological,sahni2006reconstructing,frieman2008dark,li2011dark,sotiriou2010f,de2010f,capozziello2011extended,Chevallier:2000qy,Linder:2002et,Park:2024pew, 
Nojiri:2017ncd,Piedipalumbo:2023dzg,Benetti:2019gmo,Guin:2025xki}. These alternative scenarios usually feature dynamical energy densities that effectively reproduce the behavior of $\Lambda$. To explore the variety of possible models, we adopt parameterizations of cosmological quantities at the background level. This includes models involving expansions, parameterizations or principal component analyzes of the equation of state $w\equiv{p/\rho}$ of a DE fluid with pressure $p$ and energy density $\rho$ \citep{linder1988cosmological,Chevallier:2000qy,Linder:2002et,Park:2024pew}.

In modern concordance cosmology, there are certain discrepancies attributed to data sets (for instance, the Hubble $H_0$ (about $>4\sigma$), and S8  $={\sigma_8\sqrt{\Omega_m/0.3}} $ (about 2-3 $\sigma$)  parameters measurements.) \citep{DiValentino:2020vnx,DES:2017myr,Basilakos:2017rgc,Joudaki:2017zdt}. The full-shape analyses of large-scale structure data indicate a more pronounced tension, reaching a significance level of at least $4.5\sigma$~\citep{Ivanov:2024xgb,Chen:2024vuf}, thereby suggesting a potential inconsistency between constraints derived from early and late time observations of the Universe.

Complementing these findings, early results from Planck 2013~\citep{ade2014planck}, suggested $\omega = -1.13^{+0.13}_{-0.14}$, slightly favoring the phantom regime. Subsequent improvements in supernova calibration, in 2014 Joint Light-curve Analysis (JLA) dataset~\citep{betoule2014improved} reduced this discrepancy  and combining JLA with Planck 2013 brought dark energy constraints in agreement with $\Lambda$CDM. Planck 2015~\citep{ade2016planck}, using JLA as its default supernova dataset, confirmed this consistency, yielding $\omega = -1.006^{+0.085}_{-0.091}$. However, in 2022, the Pantheon$^+$ supernova compilation~\citep{brout2022pantheon+} reported $\omega = -0.90 \pm 0.14$ (SN only), and $\omega = -0.978^{+0.024}_{-0.031}$ when combined with CMB and BAO data, consistent with $\Lambda$CDM within $2\sigma$. The Union3 compilation by \cite{rubin2025union} further supported this trend, indicating mild tension with $\Lambda$CDM at $1.7-2.6\sigma$ and favoring dynamical dark energy models with $\omega_0 > -1$ and $\omega_a < 0$.

Furthermore, in 2024, building on hints from the Pantheon$^+$ and Union3 catalogs, DESY5 found that best-fit values of $\omega$ were consistently slightly greater than $-1$ at more than the $1\sigma$ level, both for supernova data alone and combined with CMB, BAO, and $3\times 2$pt measurements, supporting a trend toward mildly dynamical dark energy. DESI’s first-year BAO data~\citep{adame2025desi} further strengthened this evidence, showing deviations from $\Lambda$CDM at $2.6\sigma$–$3.9\sigma$ when combined with CMB, Pantheon+, Union3, and DESY5 datasets, favoring a dynamical dark energy scenario with $\omega_0>-1$, $\omega_a<0$, and $\omega_\omega+\omega_a<-1$ (Quintom-B). The DESI DR2 BAO data~\citep{karim2025desi} alone excludes $\Lambda$CDM at $3.1\sigma$, and up to $4.2\sigma$ when combined with other datasets, with improved precision over DR1.

In the DESI DR2 analysis, the combination of DESI DR2 + CMB + DES-SN5Y leads to a preference for dynamical dark energy at the $4.2\sigma$ level, which is significantly stronger than that obtained using other SNe~Ia samples. This striking result has motivated several recent studies \citep{gialamas2025interpreting,efstathiou2025evolving, cortes2025desi,chaudhary2025evidence,capozziello2025dark} to investigate the origin of this preference. These works have shown that 194 SNe~Ia in the DES-SN5Y sample lie at very low redshift ($z < 0.01$) and are primarily responsible for the apparent preference for dynamical dark energy. When these low-$z$ SNe Ia are excluded from the analysis, the consistency with the standard $\Lambda$CDM model is largely restored. These results show the possibility of systematic effects in the DES-SN5Y sample. Indeed, \cite{huang2025desi} showed, using a systematic diagnosis, that there is an offset of approximately $0.043$~mag between the low-$z$ and high-$z$ subsets of the DES-SN5Y sample. The DES Collaboration recently reported improved cosmological constraints from a re-analysis of the DES-SN5Y dataset, showing that the preference for dynamical dark energy persists but with a reduced statistical significance of $3.2\sigma$, compared to the earlier $4.2\sigma$ result obtained with DES-SN5Y \citep{popovic2025dark}. These discrepancies have inspired extensive investigations into the underlying new physics and its fundamental nature, leading to the development of a wide range of models that describe dark energy as a dynamical phenomenon~\citep{choudhury2025cosmological,choudhury2025cosmology,choudhury2024updated,liu2025torsion,lee2025constraining,lee2025unveiling,mazumdar2025constraint,barua2025constraints,pedrotti2025bao,jiang2024nonparametric,vagnozzi2020new,vagnozzi2023seven,fazzari2025cosmographic}.

The persistent discrepancies observed across multiple cosmological datasets call into question the completeness of the $\Lambda$CDM model, potentially motivating refinements in modeling techniques or extensions to its theoretical foundation. A model-independent approach to probing such deviations involves the parameterization of the dark energy equation of state (EoS), $\omega(z)$. This study considers different parameterizations of the EoS, based on the Chevallier–Polarski–Linder (CPL) form~\citep{Chevallier:2000qy}, also known as parameterization $\omega_0\omega_a$. The CPL framework enables the investigation of deviations from a true cosmological constant, and when applied to combined analyses of CMB data and various Type Ia supernova datasets, it reveals a departure from $\omega = -1$.

In this paper, we go beyond the standard CPL parametrization and consider a broad and extended class of dark energy models, including $\omega$CDM, CPL, logarithmic, exponential, BA, JBP, and GDED. While several recent studies have presented cosmological constraints using DESI DR2 data in combination with external probes most notably cosmic microwave background (CMB) measurements \citep{lodha2025desi,lodha2025extended,li2025robust} the present work is deliberately designed to test the preference for dynamically evolving dark energy over the $\Lambda$CDM model using only \emph{late-time} observables. This approach allows us to test the preference for dynamical dark energy models in the absence of assumptions tied to early-Universe physics and to examine whether the preference for dynamical dark energy persists once CMB information is excluded. During our analysis, we focus on DESI DR2 BAO measurements combined with cosmic chronometers and three independent Type~Ia supernova samples (Pantheon$^+$, DES-Dovekie, and Union3). Our paper is organized as follows. In Section \ref{sec_2}, we introduce the cosmological background equations and models. Section \ref{sec_3} details the core of this work with datasets and methodology using the Markov Chain Monte Carlo (MCMC) sampling against the recent DESI DR2 dataset, while section \ref{sec_4} is dedicated to the discussion of results.  In Section \ref{sec_5}, we draw the conclusions.

\section{The Background Cosmology }\label{sec_2}
Under the two foundational conditions, i.e., spatial uniformity and directional symmetry of the cosmological principle, which has become a testable hypothesis supported by extensive observational evidence across cosmic scales (e.g., \cite{scrimgeour2012wigglez,Laurent:2016eqo}), the spacetime geometry of the Universe can be described by the Friedmann-Lema\^itre-Robertson-Walker (FLRW) spacetime metric,
\begin{equation}\label{eq_1}
ds^2 = dt^2 - a^2(t) \left[\frac{dr^2}{1 - kr^2} + r^2 \left(d\theta^2 + \sin^2\theta \, d\phi^2 \right) \right],
\end{equation}
where $r$, $\theta$, and $\phi$ denote the comoving spatial coordinates, and $t$ represents cosmic time. The parameter $k$ characterizes the spatial curvature of the three-dimensional geometry of the Universe. The evolution of the Universe is governed by the scale factor $a(t)$, which is a function of the energy densities and pressures of the components that fill the Universe. This time, evolution is formally governed by the two Friedmann equations, which follow from the Einstein field equations 
\begin{equation}\label{eq_2}
    R_{\mu\nu}-\frac{1}{2}R g_{\mu\nu}=8\pi G \left(T_{\mu\nu}-\frac{\Lambda}{8\pi G } g_{\mu\nu}\right)
\end{equation}
when applied to a FLRW spacetime:
\begin{align}
H^2 \equiv \left(\frac{\dot{a}}{a} \right)^2 &= \frac{8\pi G}{3} \rho_i - \frac{k}{a^2} + \frac{\Lambda}{3} \ , \label{eq_3} \\
\frac{\ddot{a}}{a} &= -\frac{4\pi G}{3} \left( \rho_i + 3p_i \right) + \frac{\Lambda}{3} \ , \label{eq_4}
\end{align}

Here, $H$ denotes the Hubble expansion rate, $\Lambda$ represents the cosmological constant, $\rho_i$ is the aggregate energy density of all cosmic components, and $p_i$ is the corresponding pressure. Although the cosmological constant $\Lambda$ can be absorbed into the overall energy density, it is conventionally kept explicit to honor its historical introduction in Einstein’s equations.  In light of observations that confirm cosmic acceleration, $\Lambda$ is now interpreted as the simplest form of dark energy.

Dark energy influences the expansion rate through two primary quantities: its current density parameter relative to the critical density, $\Omega_{\rm de}$, and its equation-of-state parameter, $\omega$. The most straightforward assumption is that the equation-of-state parameter remains fixed in time. However, in  the general case, \(\omega\) may vary with cosmic time or redshift. The energy–momentum conservation equation  
\begin{equation}
\dot{\rho_i} + 3H\,(\rho_i + p_i) = 0, \label{eq_5}
\end{equation}
is not independent but follows directly from combining Eqs.~\eqref{eq_3} and \eqref{eq_4}. Substituting the solution  of Eq.~\eqref{eq_5} that represents the  redshift evolution of the individual energy components into Eq.~\eqref{eq_3} yields the Hubble expansion function, which governs the dynamical evolution of the Universe:
\begin{equation}\label{eq_6}
\begin{split}
E(z) = & \Bigl[\, 
    \Omega_{r}(1+z)^4  
    + \Omega_{m}(1+z)^3 
    + \Omega_{k}(1+z)^2  \\
    & + \Omega_{\Lambda}\, f_{\mathrm{DE}}(z)
\Bigr] ,
\end{split}
\end{equation}
where $E(z) \equiv H^2(z)/H_0^2$ is the dimensionless Hubble function, $H_0$ is the present-day Hubble constant, and $\Omega_{r}$, $\Omega_{m} = \Omega_{b} + \Omega_{\mathrm{cdm}}$, $\Omega_{k}$, and $\Omega_{\Lambda}$ are the density parameters of radiation, matter (baryonic plus cold dark matter), spatial curvature, and dark energy, respectively. Also, the function $f_{\mathrm{DE}}(z)$ denotes the redshift dependence of the dark energy component and is defined as
\begin{equation}\label{eq_7}
f_{\mathrm{DE}}(\equiv \rho_{DE}(z)/\rho_{DE,0}) = \exp\!\left[3\int_0^z \frac{1+\omega(z')}{1+z'}\,dz'\right].
\end{equation}
For a cosmological constant ($\omega = -1$), $f_{\mathrm{DE}}(z) = 1$, and Eq.~\eqref{eq_6} reduces to the standard o$\Lambda$CDM,
\begin{equation}\label{eq_8}
    \begin{aligned}
    E(z) = & \left[ \Omega_{r}(1+z)^4 + \Omega_{m}(1+z)^3 + \Omega_{k}(1+z)^2 \right. \\
    & \left. + \Omega_{\Lambda} \right] ~.
    \end{aligned}
\end{equation}

In the special case of a spatially flat Universe ($\Omega_{k0}=0$), Eq.~\eqref{eq_8} reduces to
\begin{equation}\label{eq_9}
E(z) = \left[\, \Omega_{r}(1+z)^4 + \Omega_{m}(1+z)^3 + \Omega_{\Lambda} \,\right].
\end{equation}
In the case of a constant $\omega$, Eq.~\eqref{eq_7} simplifies to $(1+z)^{3(1+\omega)}$, and substituting this into Eq.~\eqref{eq_6} yields the corresponding Hubble function of the o$\omega$CDM model.

\begin{equation}\label{eq_10}
    \begin{aligned}
    E(z) = & \left[ \Omega_{r}(1+z)^4 + \Omega_{m}(1+z)^3  + \Omega_{k}(1+z)^2 \right. \\
    & \left. + \Omega_{\Lambda}(1+z)^{3(1+\omega)} \right] ~.
    \end{aligned}
\end{equation}

Taking into account $\Omega_{k0} = 0$ in Eq.~\eqref{eq_10}, it reduces to the spatially flat $\omega$CDM model.
\begin{equation}\label{eq_11}
    \begin{aligned}
    E(z) = & \left[ \Omega_{r}(1+z)^4 + \Omega_{m}(1+z)^3  +  \right. \\
    & \left. \Omega_{\Lambda}(1+z)^{3(1+\omega)} \right] ~.
    \end{aligned}
\end{equation}
We also consider several dynamical dark energy models that correspond to different functional forms of $\omega(z)$, as summarized in Table~\ref{tab_1} (2nd column). For each model, using Eq.~\eqref{eq_7}, we can derive the corresponding form of $f_{\mathrm{DE}}(z)$ (3rd column). These expressions for $f_{\mathrm{DE}}(z)$ can then be substituted in Eq.~\eqref{eq_6} to obtain the corresponding Hubble function, which characterizes the expansion history for each model.
\begin{table*}
\resizebox{\textwidth}{!}{%
\begin{tabular}{|c|c|c|c|}
\hline
\textbf{Parameterization} & \textbf{$\omega(z)$} & \textbf{$f_{DE}(z)$} & \textbf{Reference} \\
\hline
$\omega_0\omega_a$CDM & $\omega_0 + \frac{z}{1+z} \omega_a$ & $(1+z)^{3(1+\omega_0+\omega_a)} e^{-\frac{3\omega_a z}{1+z}}$ & \citep{chevallier2001accelerating,linder2003exploring} \\
Logarithmic & $\omega_0 + \omega_a \log(1+z)$ & $(1+z)^{3(1+\omega_0)} e^{\frac{3}{2} \omega_a(\log(1+z))^2}$ & \citep{efstathiou1999constraining,silva2012thermodynamics} \\
Exponential & $\omega_0+\omega_a\left(e^{\frac{z}{1+z}}-1\right)$ & $e^{\!\left[3 \omega_a\!\left(\frac{-z}{1+z}\right)\right]\,
(1+z)^{3(1+\omega_0+\omega_a)}\,}
e^{\!\left[3 \omega_a\!\left(\frac{1}{4(1+z)^2}+\frac{1}{2(1+z)}-\frac{3}{4}\right)\right]\,
(1+z)^{\tfrac32 \omega_a}}$ & \citep{najafi2024dynamical} \\
JBP & $\omega_0 + \frac{z}{(1+z)^2} \omega_a $ & $(1+z)^{3(1+\omega_0)} e^{\frac{3\omega_a z^2}{2(1+z)^2}}$ & \citep{jassal2005wmap} \\
BA & $\omega_0 + \frac{z(1+z)}{1+z^2} \omega_a$ & $(1+z)^{3(1+\omega_0)}(1+z^2)^{\frac{3\omega_a}{2}}$ & \citep{barboza2008parametric} \\
GEDE & $-1-\frac{\Delta}{3\ln (10)}\left[ 1+\tanh\left( \Delta \log_{10}\left(\frac{1+z}{1+z_t}\right)\right)\right]$ & $\left( \frac{1 - \tanh\left(\Delta \times \log_{10} \left( \frac{1+z}{1+z_t} \right)\right)}{1 + \tanh\left(\Delta \times \log_{10} (1 + z_t)\right)} \right)$ & \citep{li2020evidence,Sharma:2025qmv} \\
\hline
\end{tabular}
}
\caption{Dark energy parameterizations with their equations of state $\omega(z)$, evolution functions $f_{DE}(z)$}\label{tab_1}
\end{table*}
\section{Datasets and Methodology}\label{sec_3}
In our analysis, we use the \texttt{SimpleMC}\footnote{\protect\url{https://github.com/ja-vazquez/SimpleMC.git}} cosmological inference code to estimate the posterior distributions of parameters for each cosmological model. In this code, we use the Metropolis Hastings Markov Chain Monte Carlo (MCMC) algorithm \citep{hastings1970monte}, which allows efficient exploration of the parameter space. The convergence of Markov chains is tested using the Gelman–Rubin diagnostic ($R-1$) \citep{gelman1992inference}, applying a strict threshold of $R-1 < 0.01$. The MCMC results are subsequently analyzed and visualized using the \texttt{GetDist} package~\citep{lewis2025getdist}. In our analysis, we used multiple datasets, with a focus on late-time observations, to determine the posterior distributions of the model parameters. The details of these datasets are provided below.

\begin{itemize}
      \item \textbf{Cosmic Chronometers:} First, we consider Cosmic Chronometers (CC), massive, passively evolving galaxies with old stellar populations and negligible star formation, which allow direct measurements of the Hubble parameter $H(z)$ via the differential age technique \citep{jimenez2002constraining}. These measurements are valuable for model independent studies of the Universe’s expansion history. For our analysis, we use 15 measurements from \citep{moresco2012improved,moresco2015raising,moresco20166} instead of the 31 considered in \citep{vagnozzi2021eppur}, as only these consist of statistical and systematic part of the covariance matrix \citep{moresco2018setting,moresco2020setting}. While \cite{vagnozzi2021eppur} consider only the statistical part.
      
      \item \textbf{Baryon Acoustic Oscillations:} Next, we use recent baryon acoustic oscillation (BAO) measurements from over 14 million galaxies and quasars obtained by the Dark Energy Spectroscopic Instrument (DESI) Data Release 2 (DR2) \citep{karim2025desi}. These measurements are obtained from different tracers, such as Bright Galaxy Sample (BGS), Luminous Red Galaxies (LRG1), LRG2, LRG3+Emission Line Galaxies (ELG1), ELG2, Quasars (QSO), and Lyman-$\alpha$ forests (Ly$\alpha$); for further details, see Section 3 of \cite{karim2025desi}. Each tracer provides different BAO measurements expressed through various ratios: $D_M/r_d$, $D_H/r_d$, $D_V/r_d$, and $D_M/D_H$. Here, $D_H(z) = c/H(z)$ represents the Hubble distance, $D_M(z) = c \int_0^z \frac{dz'}{H(z')}$ represents the comoving angular diameter distance, and $D_V(z) \equiv [z D_M^2(z) D_H(z)]^{1/3}$ represents the volume-averaged distance. The $r_d$ corresponds to the sound horizon in the drag epoch, which in a flat $\Lambda$CDM model takes the value $r_d = 147.09 \pm 0.2$ Mpc \citep{aghanim2020planck}.
      \item \textbf{Type Ia supernovae :} Finally, we use three different supernova catalogs. First, we consider the Pantheon$^+$ sample \citep{brout2022pantheon+}, which comprises 1701 light curves from 1550 Type Ia Supernovae (SNe Ia). In our analysis, we exclude the light curves below \(z < 0.01\), as these light curves are significantly affected by systematic uncertainties arising from peculiar velocities. Next, we use the re-calibrated 1,820 photometric Type Ia supernova light curves obtained over five years by the Dark Energy Survey Supernova Program (DES-Dovekie)~\citep{popovic2025dark}. This catalog consists of 1,623 DES SNe~Ia, with 197 low-redshift ($z < 0.1$) SNe~Ia from the CfA3-4/CSP Foundation sample~\citep{hicken2009cfa3,hicken2012cfa4,foley2017foundation}. The revised DES-Dovekie has 1,718 SNe~Ia overlapping between DES-Dovekie and DES SN5YR~\citep{abbott2024dark}. Finally, we use the Union3 compilation \citep{rubin2025union}, which comprises 2,087 SNe Ia, including 1,363 events overlapping with the Pantheon$^+$ sample. For each sample, we marginalize the parameter $\mathcal{M}$; see Equations (A9–A12) in \cite{goliath2001supernovae} for further details.
\end{itemize}
To evaluate and compare the statistical performance of each model, we compute the Bayesian evidence $\ln \mathcal{Z}$ using the MCEvidence framework~\citep{heavens2017marginal}. This metric quantifies the goodness of fit while penalizing the complexity of the model. Model comparison is performed via the {Bayes factor: $B_{ab} = \frac{\mathcal{Z}_a}{\mathcal{Z}_b},$ or equivalently, through the difference in logarithmic evidence: $\Delta \ln \mathcal{Z} = \ln \mathcal{Z}_a - \ln \mathcal{Z}_b.$ A higher value of $\ln \mathcal{Z}$ indicates stronger statistical support for the model, while a lower value suggests weaker support. The strength of evidence is interpreted using the revised Jeffreys scale \citep{kass1995bayes}:
\begin{itemize}
    \item $|\Delta \ln \mathcal{Z}| < 1$: Inconclusive / Weak evidence,
    \item $1 \leq |\Delta \ln \mathcal{Z}| < 3$: Moderate evidence,
    \item $3 \leq |\Delta \ln \mathcal{Z}| < 5$: Strong evidence,
    \item $|\Delta \ln \mathcal{Z}| \geq 5$: Decisive evidence.
\end{itemize}
In addition to the logarithmic Bayesian evidence, we compute the difference in the minimum chi-square, $\Delta \chi^2_{\min}$, defined as $\Delta \chi^2_{\min} \equiv \chi^2_{\min,\Lambda{\rm CDM}} - \chi^2_{\min,{\rm Model}}$. While $\Delta \chi^2_{\min}$ quantifies the improvement in the goodness of fit relative to the $\Lambda$CDM model, the Bayesian evidence provides a complementary criterion for model comparison that naturally incorporates both the quality of the fit and the complexity of the model. Together, these two statistics offer a consistent and robust assessment of the preference for dynamical dark energy.

In our analysis, we adopt several assumptions. Specifically, for a dynamical dark energy model, we assume a flat scenario ($\Omega_{k} = 0$) and compute the present-day radiation density parameter as $\Omega_{r} = 2.469 \times 10^{-5} , h^{-2} \left( 1 + 0.2271 N_{\rm eff} \right)$ following \citep{komatsu2009five}, where $N_{\rm eff} = 3.04$ is the standard effective number of relativistic species \citep{mangano2002precision}. Under these assumptions, the dark energy density parameter is given by $\Omega_{\Lambda} = 1 - \Omega_{r} - \Omega_{m} - \Omega_{k}.$ In the case of the flat Universe $\Omega_{k} = 0$ this realization reduces to $\Omega_{\Lambda} = 1 - \Omega_{r} - \Omega_{m}.$ As a result, both $\Omega_{\Lambda}$ and $\Omega_{r}$ are not treated as independent parameters, since they are fully determined by the remaining parameters. The priors chosen for these models are summarized in Table~\ref{tab_2}.

\begin{table}
\centering
\begin{tabular}{lll}
\hline
\textbf{Model} & \textbf{Parameter} & \textbf{Prior} \\
\hline
\multirow{2}{*}{\(\Lambda\)CDM} 
& \( \Omega_{m0} \) & \( \mathcal{U}[0, 1] \) \\
& \( h = H_0/100 \) & \( \mathcal{U}[0, 1] \) \\
\hline
\multirow{2}{*}{o\(\Lambda\)CDM} 
& \( \Omega_{k0} \) & \( \mathcal{U}[-1, 1] \) \\
& \( \Omega_{m0},\ h \) & \( \mathcal{U}[0, 1] \) \\
\hline
\multirow{2}{*}{\(\omega\)CDM} 
& \( \omega_0 \) & \( \mathcal{U}[-3, 1] \) \\
& \( \Omega_{m0},\ h \) & \( \mathcal{U}[0, 1] \) \\
\hline
\multirow{3}{*}{o\(\omega\)CDM} 
& \( \omega_0 \) & \( \mathcal{U}[-3, 1] \) \\
& \( \Omega_{k0} \) & \( \mathcal{U}[-1, 1] \) \\
& \( \Omega_{m0},\ h \) & \( \mathcal{U}[0, 1] \) \\
\hline
\multirow{3}{*}{$\omega_0\omega_a$CDM} 
& \( \omega_0 \) & \( \mathcal{U}[-3, 1] \) \\
& \( \omega_a \) & \( \mathcal{U}[-3, 2] \) \\
& \( \Omega_{m0},\ h \) & \( \mathcal{U}[0, 1] \) \\
\hline
\multirow{3}{*}{Logarithmic} 
& \( \omega_0 \) & \( \mathcal{U}[-3, 1] \) \\
& \( \omega_a \) & \( \mathcal{U}[-3, 2] \) \\
& \( \Omega_{m0},\ h \) & \( \mathcal{U}[0, 1] \) \\
\hline
\multirow{3}{*}{Exponential} 
& \( \omega_0 \) & \( \mathcal{U}[-3, 1] \) \\
& \( \omega_a \) & \( \mathcal{U}[-3, 2] \) \\
& \( \Omega_{m0},\ h \) & \( \mathcal{U}[0, 1] \) \\
\hline
\multirow{3}{*}{JBP} 
& \( \omega_0 \) & \( \mathcal{U}[-3, 1] \) \\
& \( \omega_a \) & \( \mathcal{U}[-3, 2] \) \\
& \( \Omega_{m0},\ h \) & \( \mathcal{U}[0, 1] \) \\
\hline
\multirow{3}{*}{BA} 
& \( \omega_0 \) & \( \mathcal{U}[-3, 1] \) \\
& \( \omega_a \) & \( \mathcal{U}[-3, 2] \) \\
& \( \Omega_{m0},\ h \) & \( \mathcal{U}[0, 1] \) \\
\hline
\multirow{2}{*}{GEDE} 
& \( \Delta \) & \( \mathcal{U}[-10, 10] \) \\
& \( \Omega_{m0},\ h \) & \( \mathcal{U}[0, 1] \) \\
\hline
\end{tabular}
\caption{The parameters and their priors including uniform priors $\mathcal{U}$ and the reduced Hubble constant $h \equiv H_0/100$.}\label{tab_2}
\end{table}

\begin{table*}[t]
\centering
\setlength{\tabcolsep}{5pt}
\resizebox{\textwidth}{!}{%
\begin{tabular}{lcccccccccccc}
\hline
\textbf{Dataset/Models} & $h$ & $\Omega_m$ & $\Omega_k$ & $\omega$ or $\omega_0$ & $\omega_a$ & $\Delta$ & $|\Delta\ln\mathcal{Z}_{\Lambda\mathrm{CDM},\mathrm{Model}}|$ & $\Delta \chi^2_{\min}$ \\
\hline
\textbf{$\Lambda$CDM} \\
CC + DESI DR2 & $0.692{\pm0.010}$ & $0.297{\pm0.008}$ & --- & --- & --- & --- & 0 & 0 \\
CC + DESI DR2 + Pantheon$^{+}$ & $0.691{\pm0.010}$ & $0.304{\pm0.008}$ & --- & --- & --- & --- & 0 & 0 \\
CC + DESI DR2 + DES-Dovekie & $0.691{\pm0.010}$ & $0.305{\pm0.007}$ & --- & --- & --- & --- & 0 & 0 \\
CC + DESI DR2 + Union3 & $0.691{\pm0.010}$ & $0.303{\pm0.008}$ & --- & --- & --- & --- & 0 & 0 \\
\hline
\textbf{o$\Lambda$CDM} \\
CC + DESI DR2 & $0.683{\pm0.010}$ & $0.293{\pm0.012}$ & $0.027{\pm0.041}$ & --- & --- & --- & 2.14 & 0.18 \\
CC + DESI DR2 + Pantheon$^{+}$ & $0.675{\pm0.010}$ & $0.292{\pm0.012}$ & $0.051{\pm0.036}$ & --- & --- & --- & 1.53 & 0.88 \\
CC + DESI DR2 + DES-Dovekie & $0.674{\pm0.016}$ & $0.294{\pm0.012}$ & $0.053{\pm0.037}$ & --- & --- & --- & 1.36 & 1.04 \\
CC + DESI DR2 + Union3 & $0.676{\pm0.010}$ & $0.294{\pm0.011}$ & $0.050{\pm0.039}$ & --- & --- & --- & 1.51 & 0.86 \\
\hline
\textbf{$\omega$CDM} \\
CC + DESI DR2 & $0.675{\pm0.034}$ & $0.296{\pm0.009}$ & --- & $-0.951^{+0.110}_{-0.091}$ & --- & --- & 1.17 & 0.20 \\
CC + DESI DR2 + Pantheon$^{+}$ & $0.667{\pm0.017}$ & $0.298{\pm0.008}$ & --- & $-0.923{\pm0.044}$ & --- & --- & 0.47 & 1.74 \\
CC + DESI DR2 + DES-Dovekie & $0.665{\pm0.016}$ & $0.298{\pm0.009}$ & --- & $-0.917{\pm0.039}$ & --- & --- & 0.02 & 2.28 \\
CC + DESI DR2 + Union3 & $0.648{\pm0.022}$ & $0.298{\pm0.009}$ & --- & $-0.867{\pm0.058}$ & --- & --- & 0.96 & 2.87 \\
\hline
\textbf{o$\omega$CDM} \\
CC + DESI DR2 & $0.677^{+0.033}_{-0.040}$ & $0.296^{+0.014}_{-0.011}$ & $0.049^{+0.066}_{-0.078}$ & $-1.09^{+0.28}_{-0.13}$ & --- & --- & 2.70 & 0.19 \\
CC + DESI DR2 + Pantheon$^{+}$ & $0.660{\pm0.026}$ & $0.301{\pm0.011}$ & $0.019^{+0.049}_{-0.056}$ & $-0.937^{+0.064}_{-0.056}$ & --- & --- & 2.48 & 1.73 \\
CC + DESI DR2 + DES-Dovekie & $0.663{\pm0.016}$ & $0.297{\pm0.012}$ & $0.002\pm{0.049}$ & $-0.918^{+0.055}_{-0.046}$ & --- & --- & 2.18 & 2.28  \\
CC + DESI DR2 + Union3 & $0.649{\pm0.028}$ & $0.298{\pm0.012}$ & $-0.002^{+0.045}_{-0.058}$ & $-0.865^{+0.076}_{-0.062}$ & --- & --- & 1.05 & 2.92 \\
\hline

\textbf{$\omega_0\omega_a$CDM} \\
CC + DESI DR2 & $0.652^{+0.032}_{-0.041}$ & $0.337^{+0.048}_{-0.035}$ & --- & $-0.61^{+0.37}_{-0.32}$ & $-1.20^{+1.0}_{-1.3}$ & --- & 0.94 & 1.44 \\
CC + DESI DR2 + Pantheon$^{+}$ & $0.676^{+0.025}_{-0.021}$ & $0.306^{+0.020}_{-0.013}$ & --- & $-0.889^{+0.059}_{-0.068}$ & $-0.31^{+0.51}_{-0.43}$ & --- & 0.25 & 1.85 \\
CC + DESI DR2 + DES-Dovekie & $0.683^{+0.023}_{-0.019}$ & $0.315^{+0.017}_{-0.013}$ & --- & $-0.838^{+0.068}_{-0.086}$ & $-0.65^{+0.59}_{-0.48}$ & --- & 0.93 & 3.02 \\
CC + DESI DR2 + Union3 & $0.668{\pm0.022}$ & $0.330^{+0.019}_{-0.014}$ & --- & $-0.70{\pm0.11}$ & $-1.07{\pm0.58}$ & --- & 3.02 & 4.88 \\
\hline
\textbf{Logarithmic} \\
CC + DESI DR2 & $0.647^{+0.031}_{-0.038}$ & $0.354^{+0.042}_{-0.033}$ & --- & $-0.52{\pm0.29}$ & $-1.29^{+0.78}_{-0.88}$ & --- & 0.96 & 1.47 \\
CC + DESI DR2 + Pantheon$^{+}$ & $0.676^{+0.025}_{-0.022}$ & $0.306^{+0.020}_{-0.013}$ & --- & $-0.893^{+0.052}_{-0.059}$ & $-0.27^{+0.39}_{-0.31}$ & --- & 1.03 & 1.92 \\
CC + DESI DR2 + DES-Dovekie & $0.684^{+0.023}_{-0.018}$ & $0.316^{+0.018}_{-0.011}$ & --- & $-0.855^{+0.056}_{-0.064}$ & $-0.50{\pm 0.40}$ & --- & 0.63 & 3.03 \\
CC + DESI DR2 + Union3 & $0.668{\pm0.023}$ & $0.331^{+0.020}_{-0.015}$ & --- & $-0.729^{+0.091}_{-0.11}$ & $-0.80^{+0.48}_{-0.43}$ & --- & 2.92 & 4.57 \\
\hline
\textbf{Exponential} \\
CC + DESI DR2 & $0.629^{+0.033}_{-0.041}$ & $0.375{\pm0.046}$ & --- & $-0.77{\pm0.14}$ & $-1.07{\pm0.63}$ & --- & 0.51 & 1.46 \\
CC + DESI DR2 + Pantheon$^{+}$ & $0.675^{+0.024}_{-0.020}$ & $0.305^{+0.017}_{-0.012}$ & --- & $-0.955{\pm0.061}$ & $-0.13^{+0.20}_{-0.18}$ & --- & 1.03 & 1.85 \\
CC + DESI DR2 + DES-Dovekie & $0.684^{+0.021}_{-0.019}$ & $0.316^{+0.016}_{-0.012}$ & --- & $-0.978{\pm0.058}$ & $-0.30_{-0.21}^{+0.24}$ & --- & 0.13 & 3.04 \\
CC + DESI DR2 + Union3 & $0.668{\pm0.021}$ & $0.331^{+0.018}_{-0.015}$ & --- & $-0.938{\pm0.063}$ & $-0.51^{+0.29}_{-0.26}$ & --- & 1.85 & 4.52 \\
\hline
\textbf{JBP} \\
CC + DESI DR2 & $0.659^{+0.035}_{-0.040}$ & $0.314^{+0.025}_{-0.018}$ & --- & $-0.72^{+0.33}_{-0.20}$ & $-1.66^{+0.74}_{-1.6}$ & --- & 0.77 & 1.11 \\
CC + DESI DR2 + Pantheon$^{+}$ & $0.671{\pm0.021}$ & $0.302^{+0.014}_{-0.012}$ & --- & $-0.888{\pm0.087}$ & $-0.36^{+0.78}_{-0.69}$ & --- & 0.18 & 1.78 \\
CC + DESI DR2 + DES-Dovekie & $0.680{\pm0.019}$ & $0.312^{+0.014}_{-0.012}$ & --- & $-0.799{\pm0.098}$ & $-1.08{\pm0.80}$ & --- & 1.51 & 3.07 \\
CC + DESI DR2 + Union3 & $0.658{\pm0.021}$ & $0.320^{+0.015}_{-0.013}$ & --- & $-0.66^{+0.14}_{-0.11}$ & $-1.56^{+0.63}_{-1.0}$ & --- & 3.41 & 4.50 \\
\hline
\textbf{BA} \\
CC + DESI DR2 & $0.638^{+0.033}_{-0.040}$ & $0.363{\pm0.040}$ & --- & $-0.46^{+0.29}_{-0.34}$ & $-0.99^{+0.64}_{-0.54}$ & --- & 0.54 & 1.49\\
CC + DESI DR2 + Pantheon$^{+}$ & $0.678^{+0.024}_{-0.021}$ & $0.308^{+0.017}_{-0.013}$ & --- & $-0.898{\pm0.056}$ & $-0.19^{+0.26}_{-0.21}$ & --- & 0.70 & 1.94 \\
CC + DESI DR2 + DES-Dovekie & $0.684^{+0.022}_{-0.019}$ & $0.315^{+0.016}_{-0.012}$ & --- & $-0.865_{-0.061}^{+0.050}$ & $-0.32^{+0.27}_{-0.22}$ & --- & 0.16 & 2.98 \\
CC + DESI DR2 + Union3 & $0.669{\pm0.022}$ & $0.330^{+0.018}_{-0.015}$ & --- & $-0.740{\pm0.095}$ & $-0.54^{+0.32}_{-0.28}$ & --- & 2.62 & 4.53 \\
\hline
\textbf{GEDE} \\
CC + DESI DR2 & $0.677{\pm0.036}$ & $0.298{\pm0.009}$ & --- & --- & --- & $-0.20^{+0.60}_{-0.86}$ & 0.70 & 0.21 \\
CC + DESI DR2 + Pantheon$^{+}$ & $0.667{\pm0.016}$ & $0.298{\pm0.008}$ & --- & --- & --- & $-0.46^{+0.24}_{-0.27}$ & 1.28 & 1.76 \\
CC + DESI DR2 + DES-Dovekie & $0.667{\pm0.015}$ & $0.299{\pm0.008}$ & --- & --- & --- & $-0.50^{+0.20}_{-0.25}$ & 1.83 & 2.34 \\
CC + DESI DR2 + Union3 & $0.647{\pm0.036}$ & $0.300{\pm0.008}$ & --- & --- & --- & $-0.84^{+0.28}_{-0.34}$ & 2.80 & 3.05 \\
\hline
\end{tabular}
}
\caption{This table presents the numerical constraints on the o$\Lambda$CDM, $\omega$CDM, o$\omega$CDM, $\omega_0\omega_a$CDM, Logarithmic, Exponential, JBP, BA, and GEDE models at the 68\% ($1\sigma$) confidence level, obtained using combinations of DESI DR2 with CC measurements and different SNe~Ia catalogs (Pantheon$^{+}$, DES-Dovekie, and Union3).}\label{tab_3}
\end{table*}


\begin{figure*}
\begin{subfigure}{.3\textwidth}
\includegraphics[width=\linewidth]{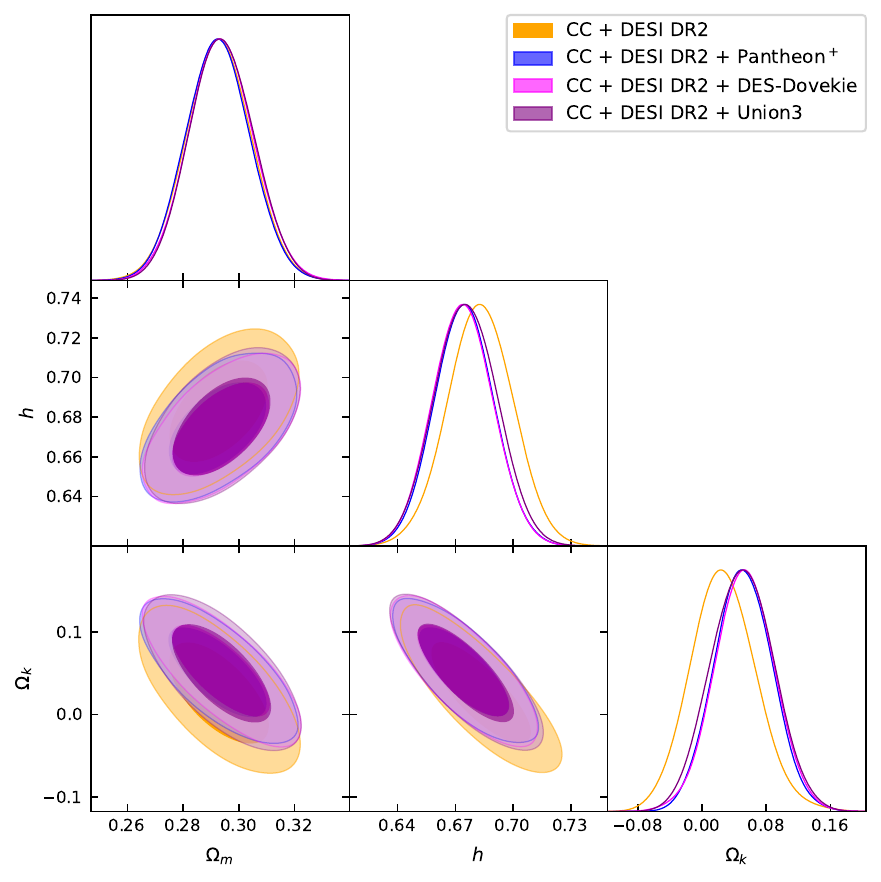}
    \caption{o$\Lambda$CDM}\label{fig_1a}
\end{subfigure}
\hfil
\begin{subfigure}{.3\textwidth}
\includegraphics[width=\linewidth]{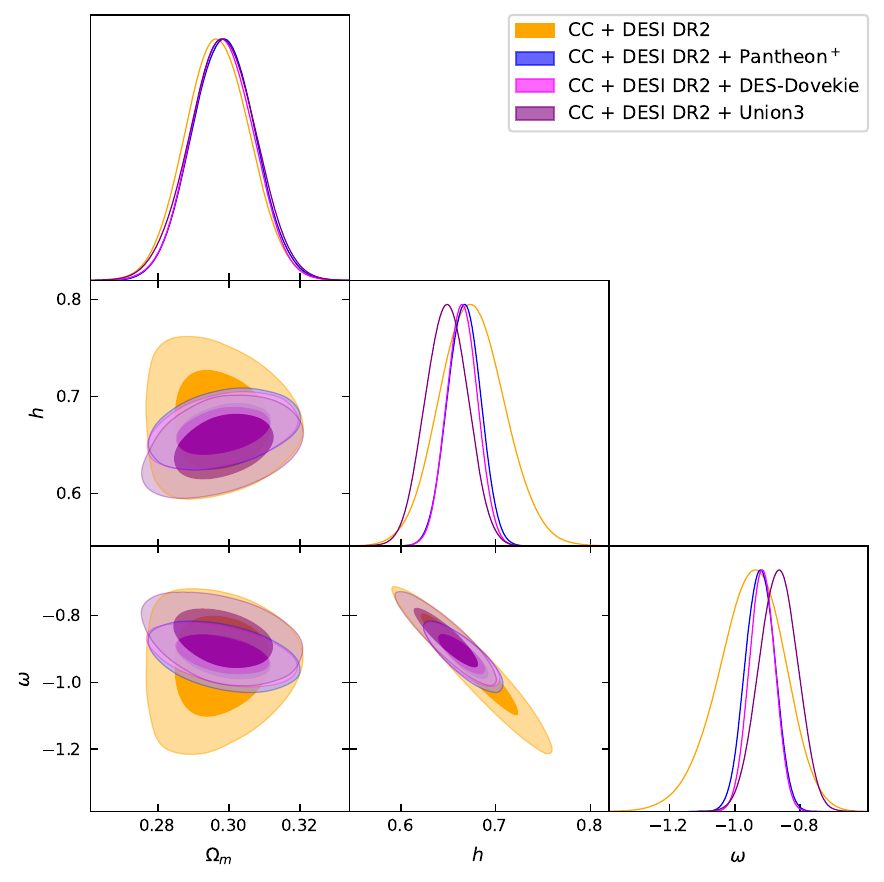}
    \caption{$\omega$CDM}\label{fig_1b}
\end{subfigure}
\hfil
\begin{subfigure}{.3\textwidth}
\includegraphics[width=\linewidth]{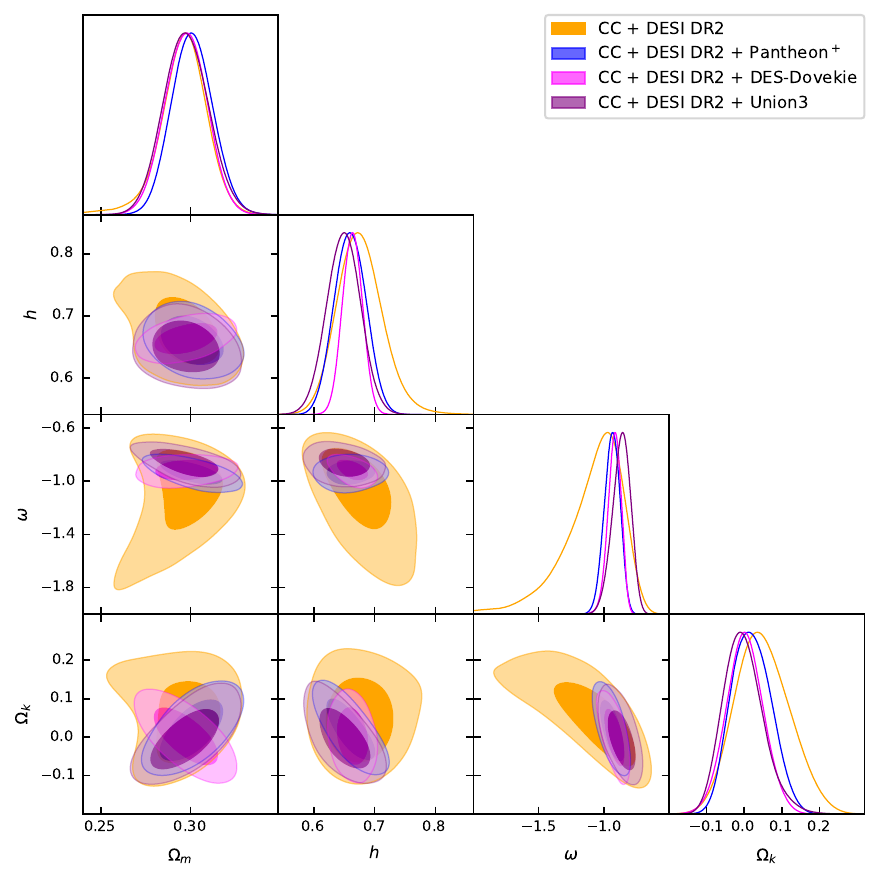}
    \caption{o$\omega$CDM}\label{fig_1c}
\end{subfigure}
\begin{subfigure}{.3\textwidth}
\includegraphics[width=\linewidth]{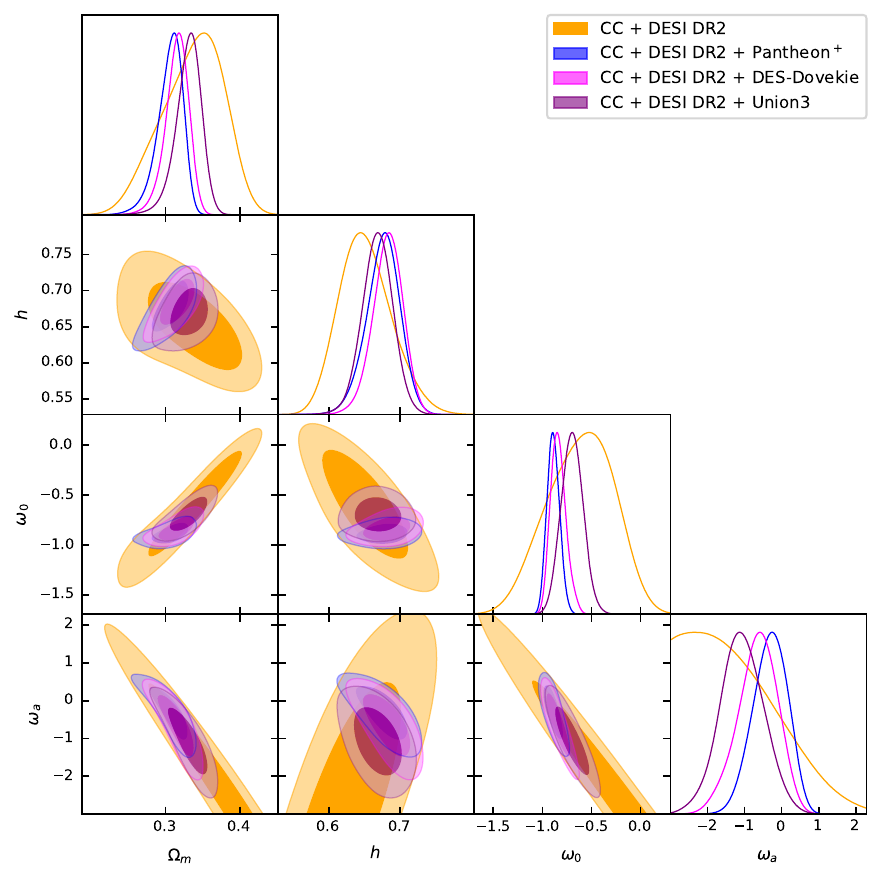}
    \caption{$\omega_0\omega_a$CDM}\label{fig_1d}
\end{subfigure}
\hfil
\begin{subfigure}{.3\textwidth}
\includegraphics[width=\linewidth]{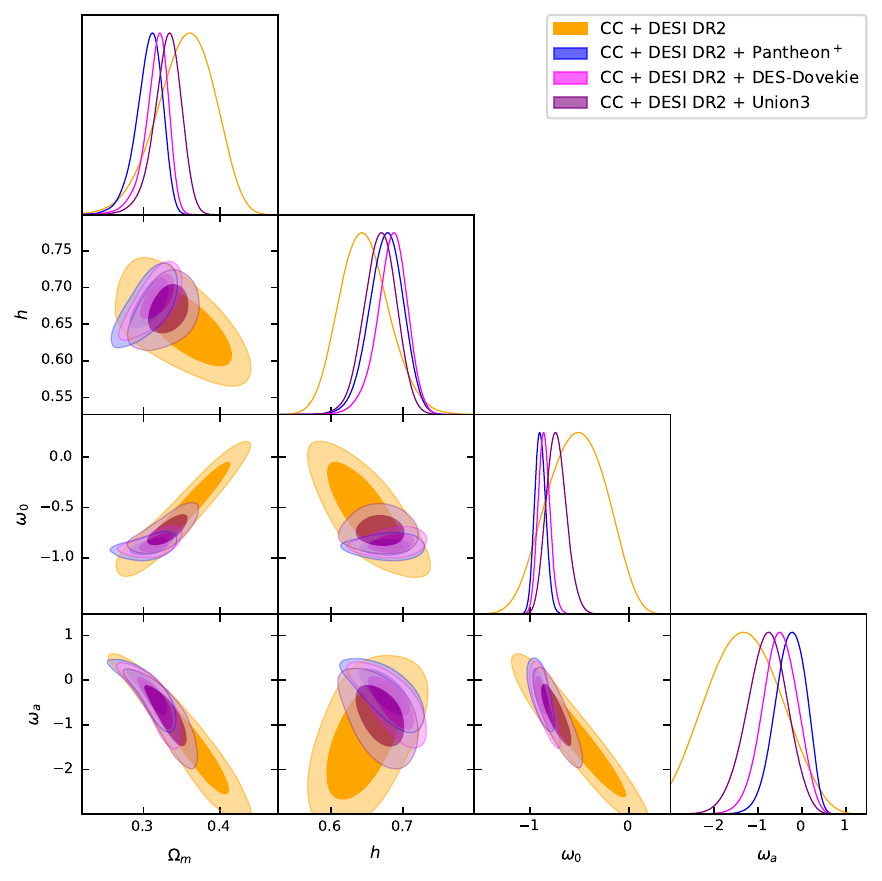}
     \caption{Logarithmic}\label{fig_1e}
\end{subfigure}
\hfil
\begin{subfigure}{.3\textwidth}
\includegraphics[width=\linewidth]{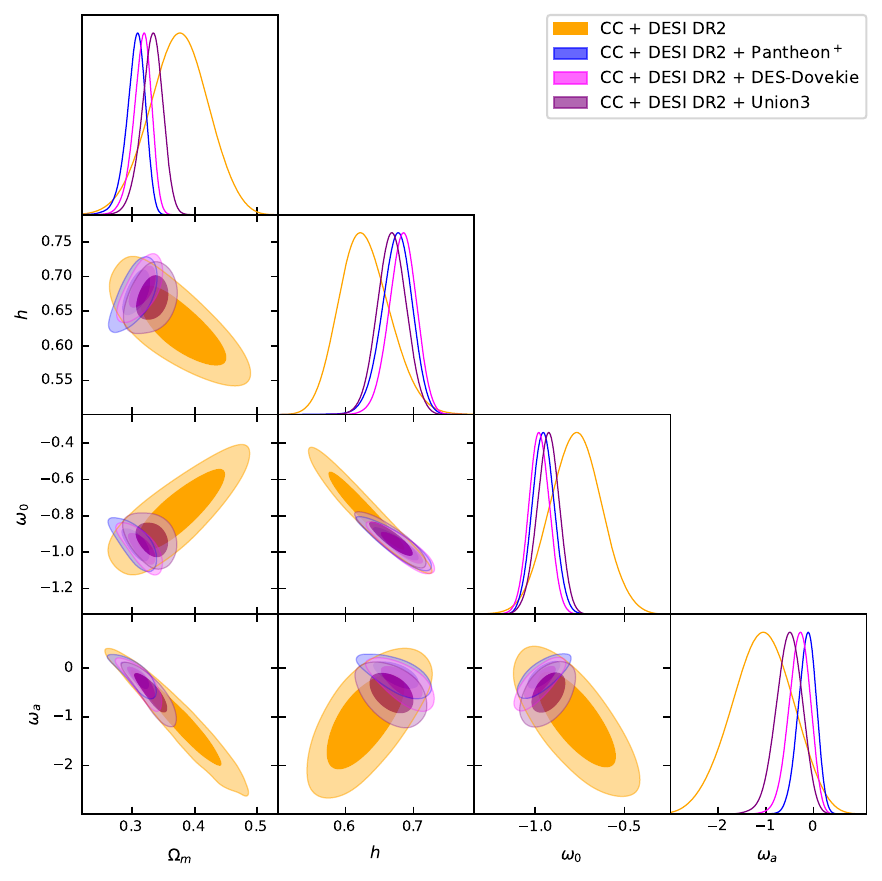}
    \caption{Exponential}\label{fig_1f}
\end{subfigure}
\hfil
\begin{subfigure}{.3\textwidth}
\includegraphics[width=\linewidth]{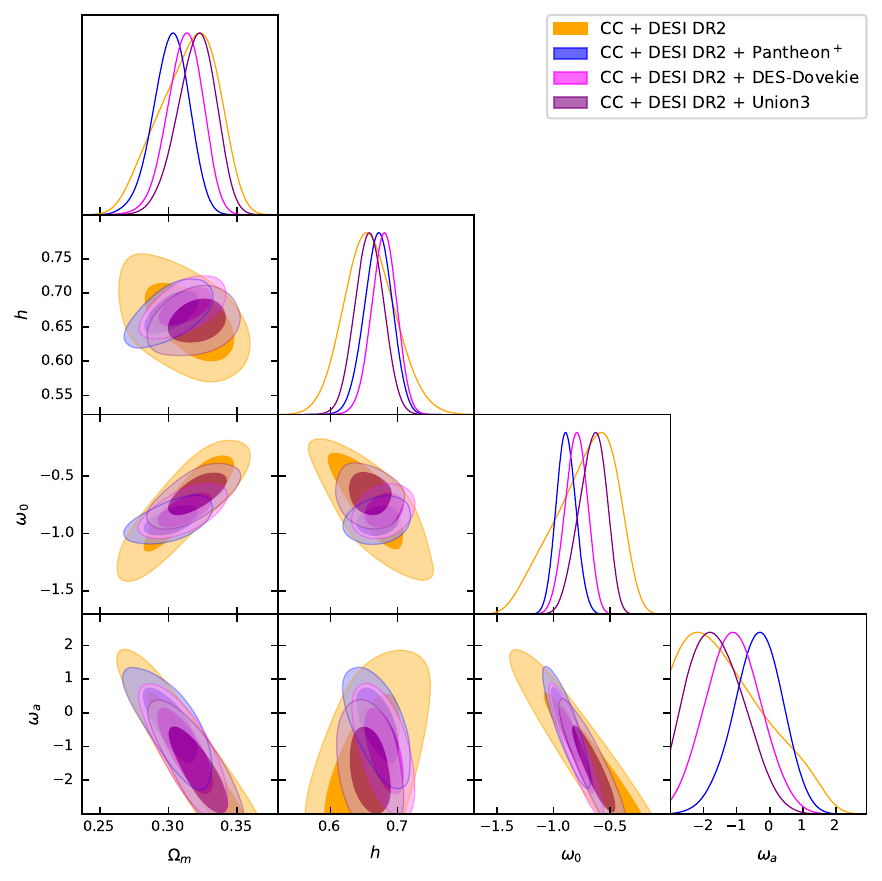}
    \caption{JBP}\label{fig_1g}
\end{subfigure}
\hfil
\begin{subfigure}{.3\textwidth}
\includegraphics[width=\linewidth]{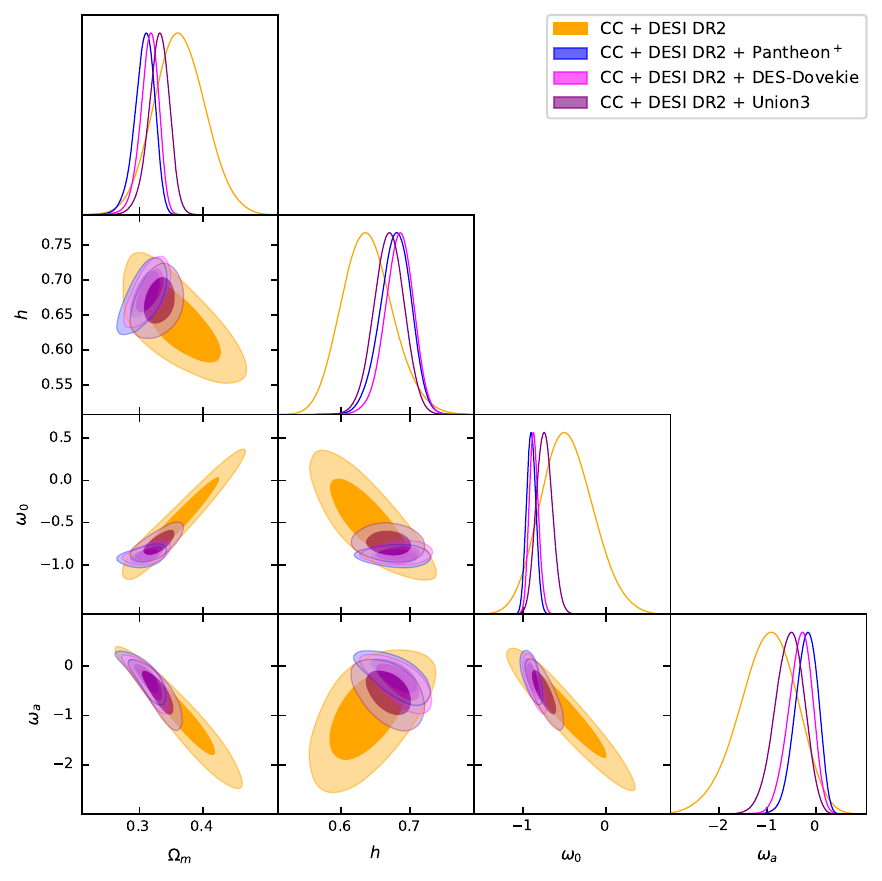}
     \caption{BA}\label{fig_1h}
\end{subfigure}
\hfil
\begin{subfigure}{.3\textwidth}
\includegraphics[width=\linewidth]{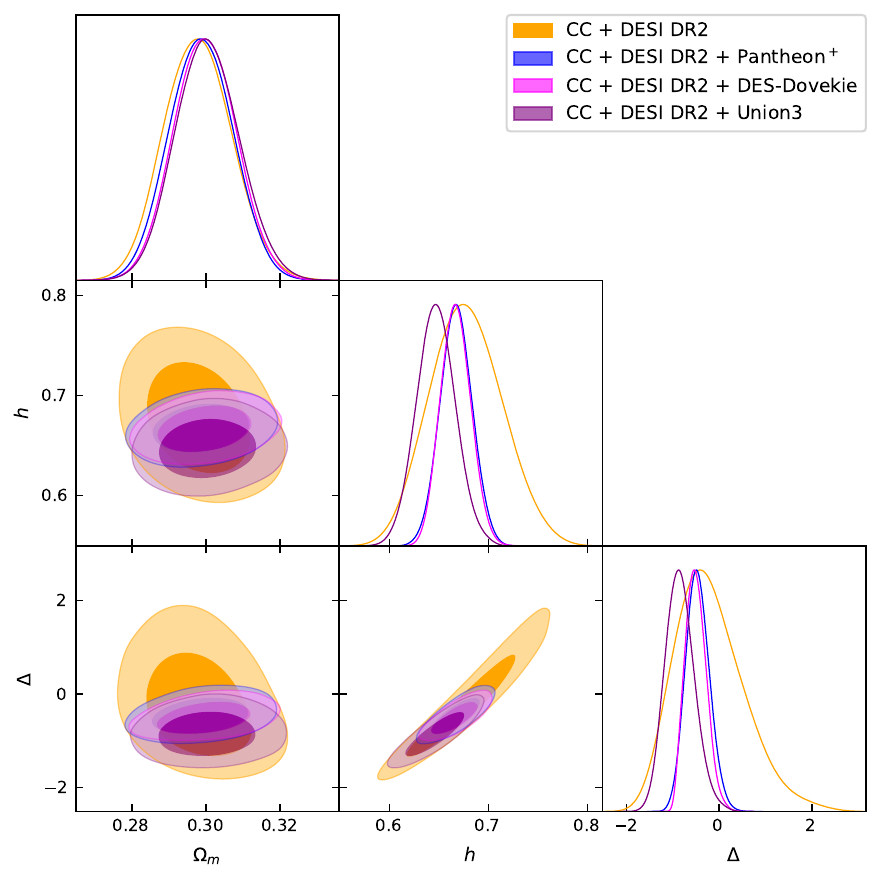}
     \caption{GEDE}\label{fig_1i}
\end{subfigure}
\caption{The figure shows the corner plot of the o$\Lambda$CDM, $\omega$CDM, o$\omega$CDM, $\omega_a \omega_0$CDM, Logarithmic, Exponential, JBP, BA, and GEDE models obtained using DESI DR2 with CC measurements and different SNe~Ia catalogs (Pantheon$^{+}$, DES-Dovekie, and Union3), at the 68\% ($1\sigma$) and 95\% ($2\sigma$) confidence levels.}\label{fig_1}
\end{figure*}

\begin{figure*}
\begin{subfigure}{.3\textwidth}
\includegraphics[width=\linewidth]{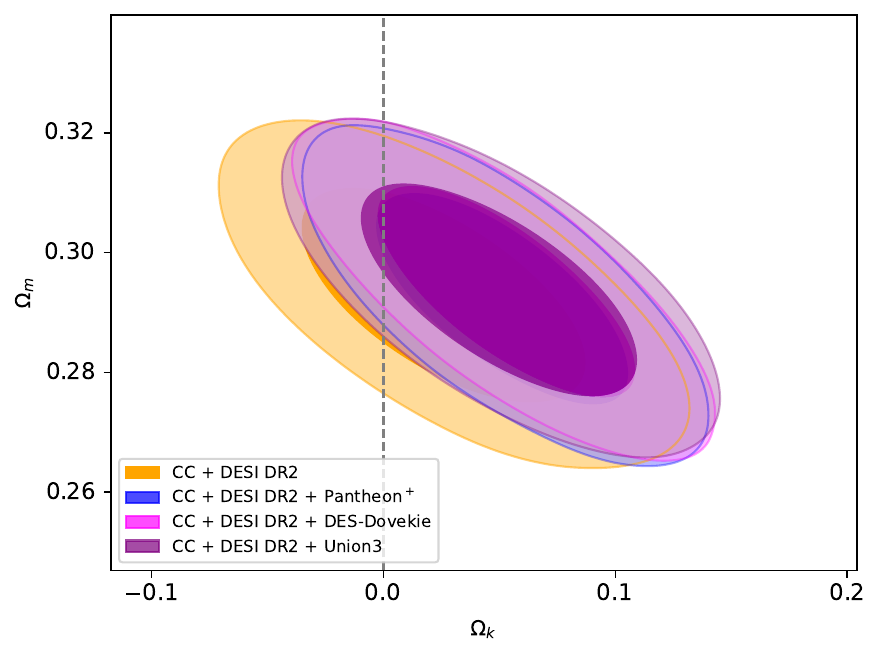}
    \caption{o$\Lambda$CDM}\label{fig_2a}
\end{subfigure}
\hfil
\begin{subfigure}{.3\textwidth}
\includegraphics[width=\linewidth]{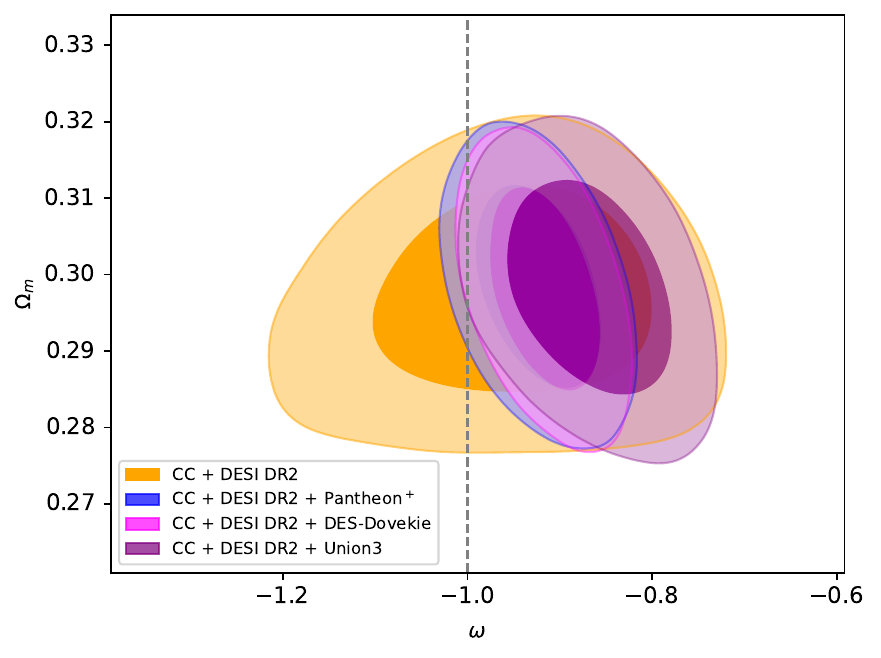}
    \caption{$\omega$CDM}\label{fig_2b}
\end{subfigure}
\hfil
\begin{subfigure}{.3\textwidth}
\includegraphics[width=\linewidth]{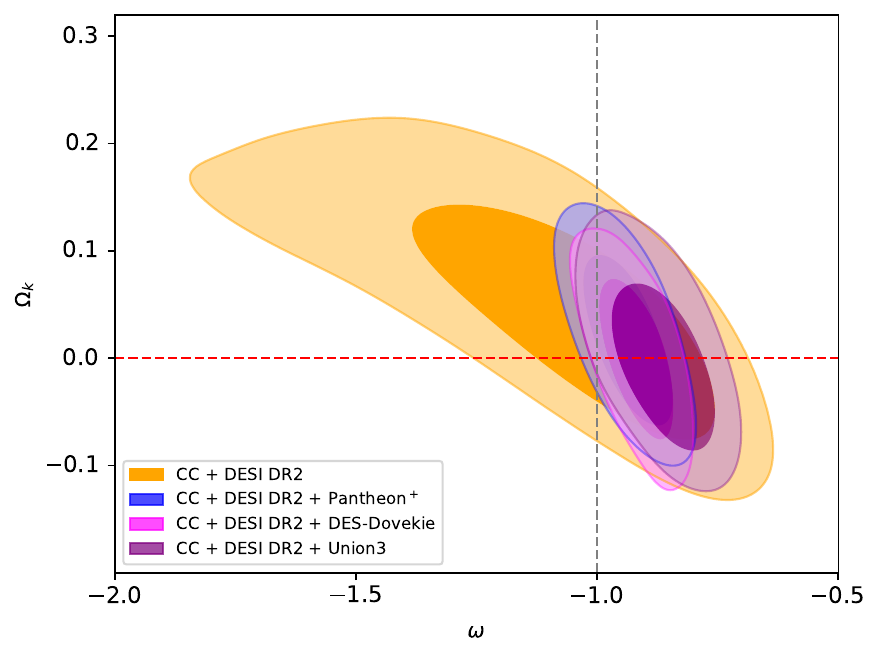}
    \caption{o$\omega$CDM}\label{fig_2c}
\end{subfigure}
\begin{subfigure}{.3\textwidth}
\includegraphics[width=\linewidth]{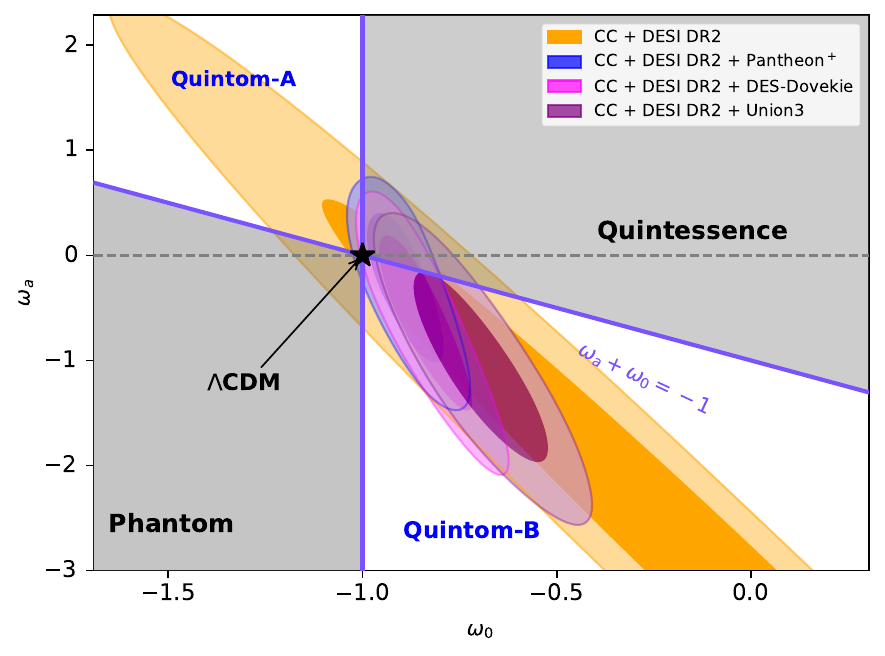}
    \caption{$\omega_0\omega_a$CDM}\label{fig_2d}
\end{subfigure}
\hfil
\begin{subfigure}{.3\textwidth}
\includegraphics[width=\linewidth]{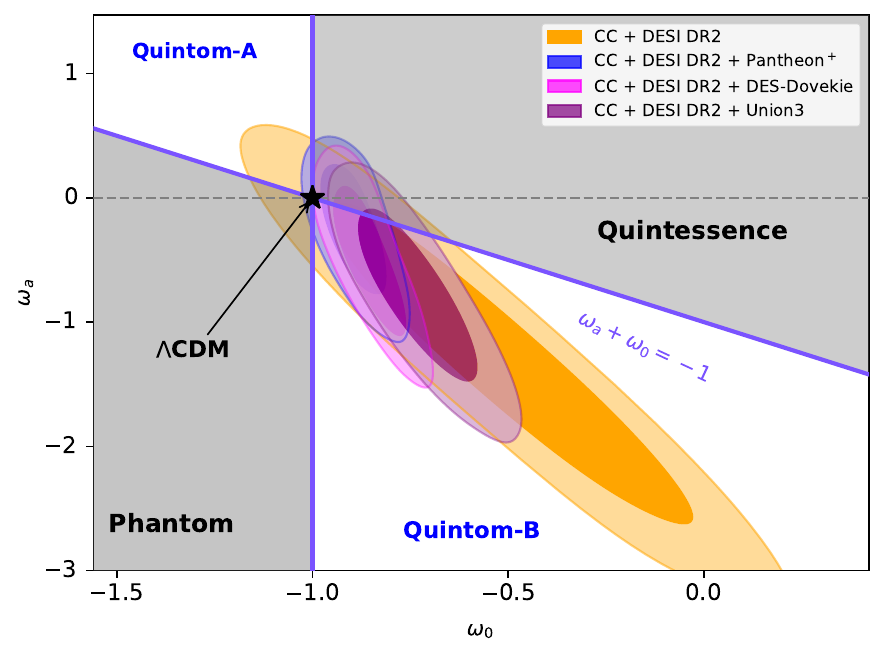}
     \caption{Logarithmic}\label{fig_2e}
\end{subfigure}
\hfil
\begin{subfigure}{.3\textwidth}
\includegraphics[width=\linewidth]{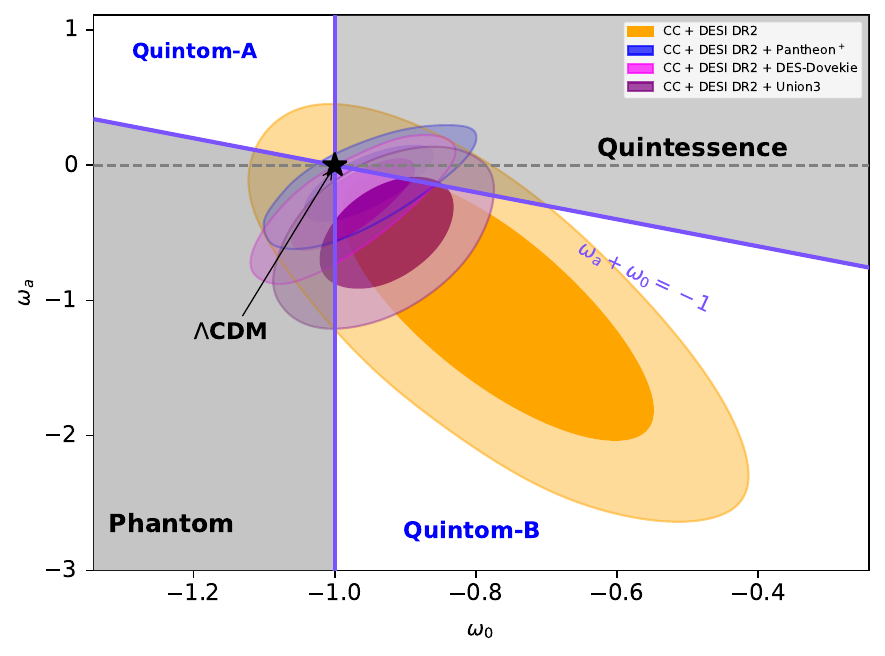}
    \caption{Exponential}\label{fig_2f}
\end{subfigure}
\hfil
\begin{subfigure}{.3\textwidth}
\includegraphics[width=\linewidth]{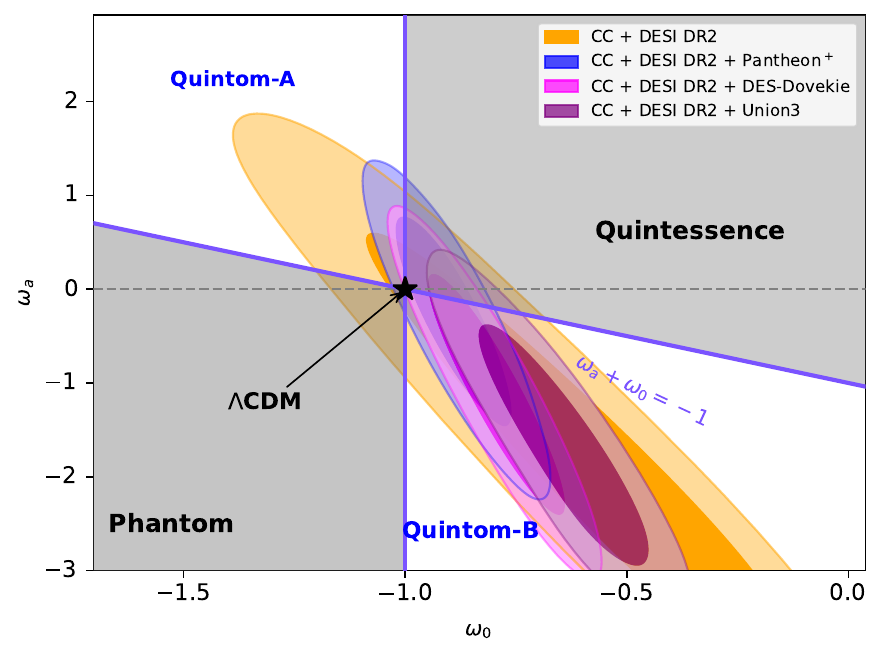}
    \caption{JBP}\label{fig_2g}
\end{subfigure}
\hfil
\begin{subfigure}{.3\textwidth}
\includegraphics[width=\linewidth]{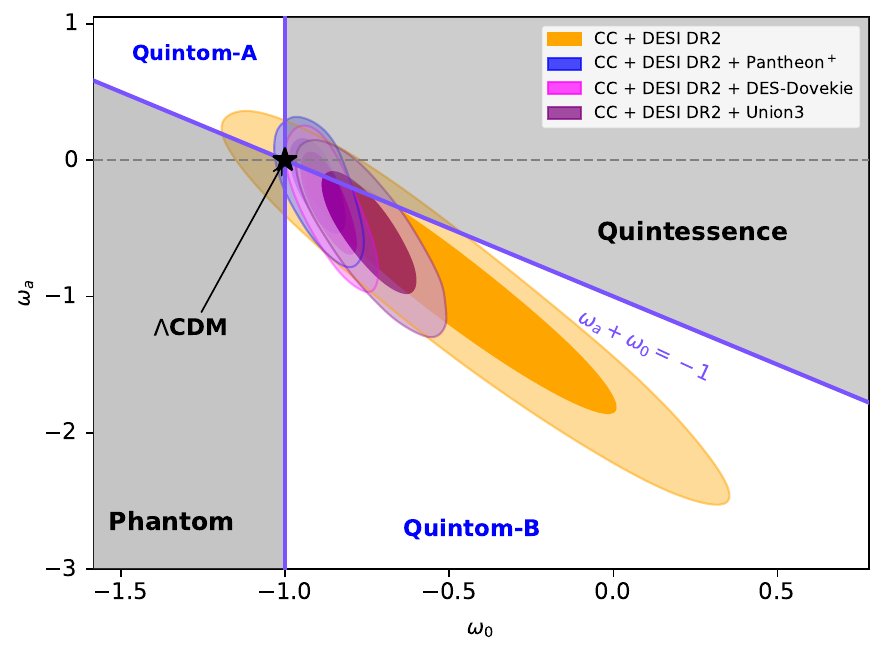}
     \caption{BA}\label{fig_2h}
\end{subfigure}
\hfil
\begin{subfigure}{.3\textwidth}
\includegraphics[width=\linewidth]{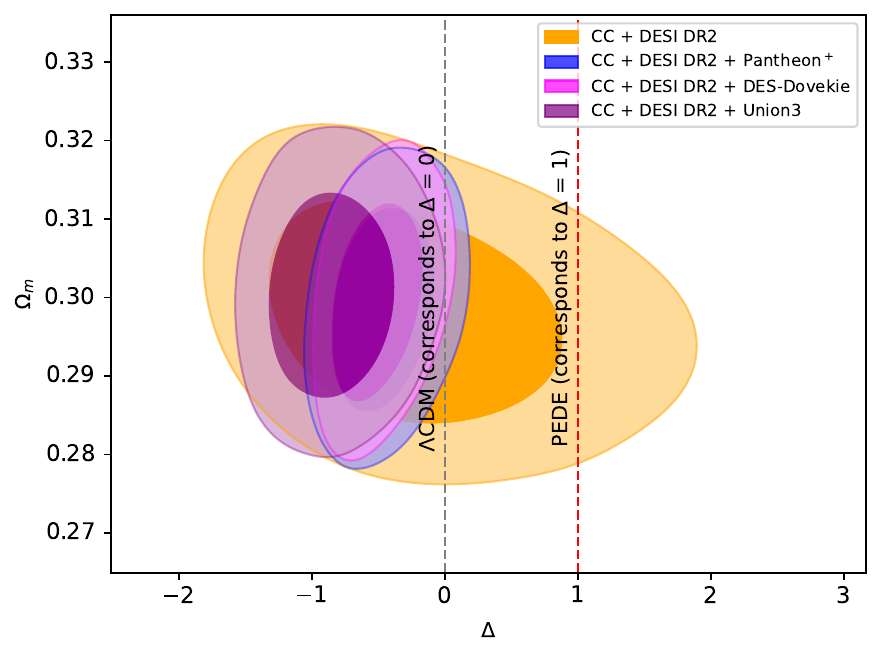}
     \caption{GEDE}\label{fig_2i}
\end{subfigure}
\caption{The figure shows the posterior distributions of different planes of the o$\Lambda$CDM , $\omega$CDM, o$\omega$CDM, $\omega_a \omega_0$CDM, Logarithmic, Exponential, JBP, BA, and GEDE models obtained using DESI DR2 with CC measurements and different SNe~Ia catalogs (Pantheon$^{+}$, DES-Dovekie, and Union3), at the 68\% ($1\sigma$) and 95\% ($2\sigma$) confidence levels.}\label{fig_2}
\end{figure*}

\begin{figure}
\centering
\includegraphics[scale=0.6]{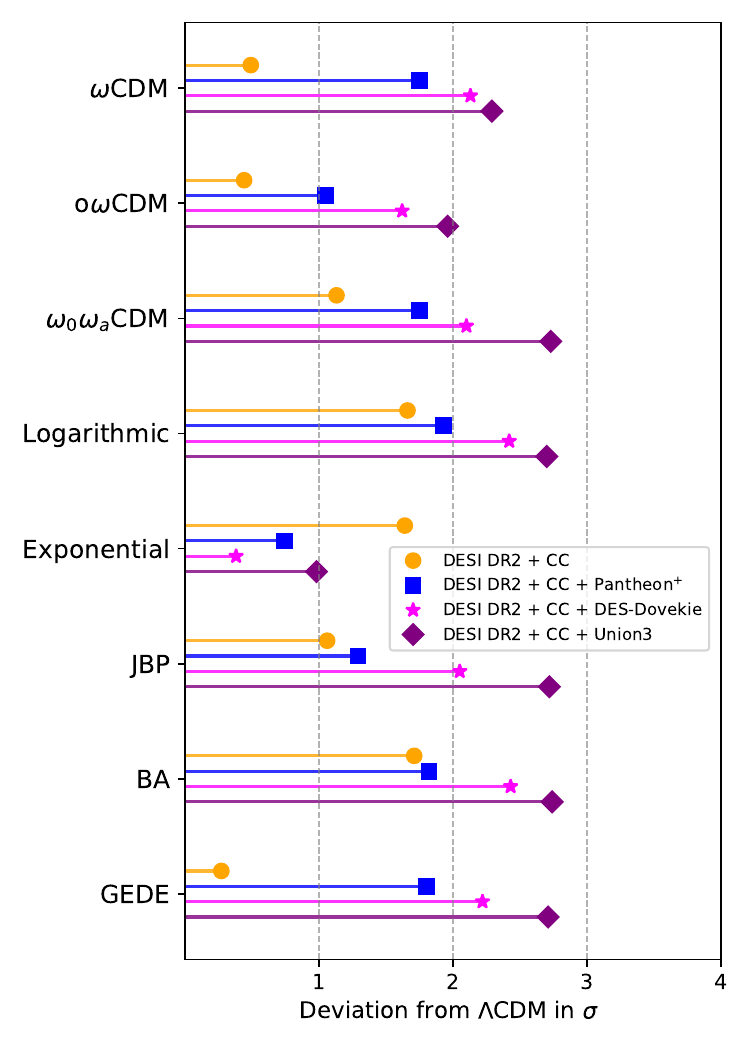}
\caption{This figure shows the statistical deviation (in units of $\sigma$) of each dark-energy model relative to the $\Lambda$CDM model, defined by $\omega = -1$ (and $\Delta = 0$ for the GEDE model). These results are obtained using DESI DR2 with CC measurements and different SNe~Ia catalogs (Pantheon$^{+}$, DES-Dovekie, and Union3).}\label{fig_3}
\end{figure}

\begin{figure*}
\centering
\includegraphics[scale=0.65]{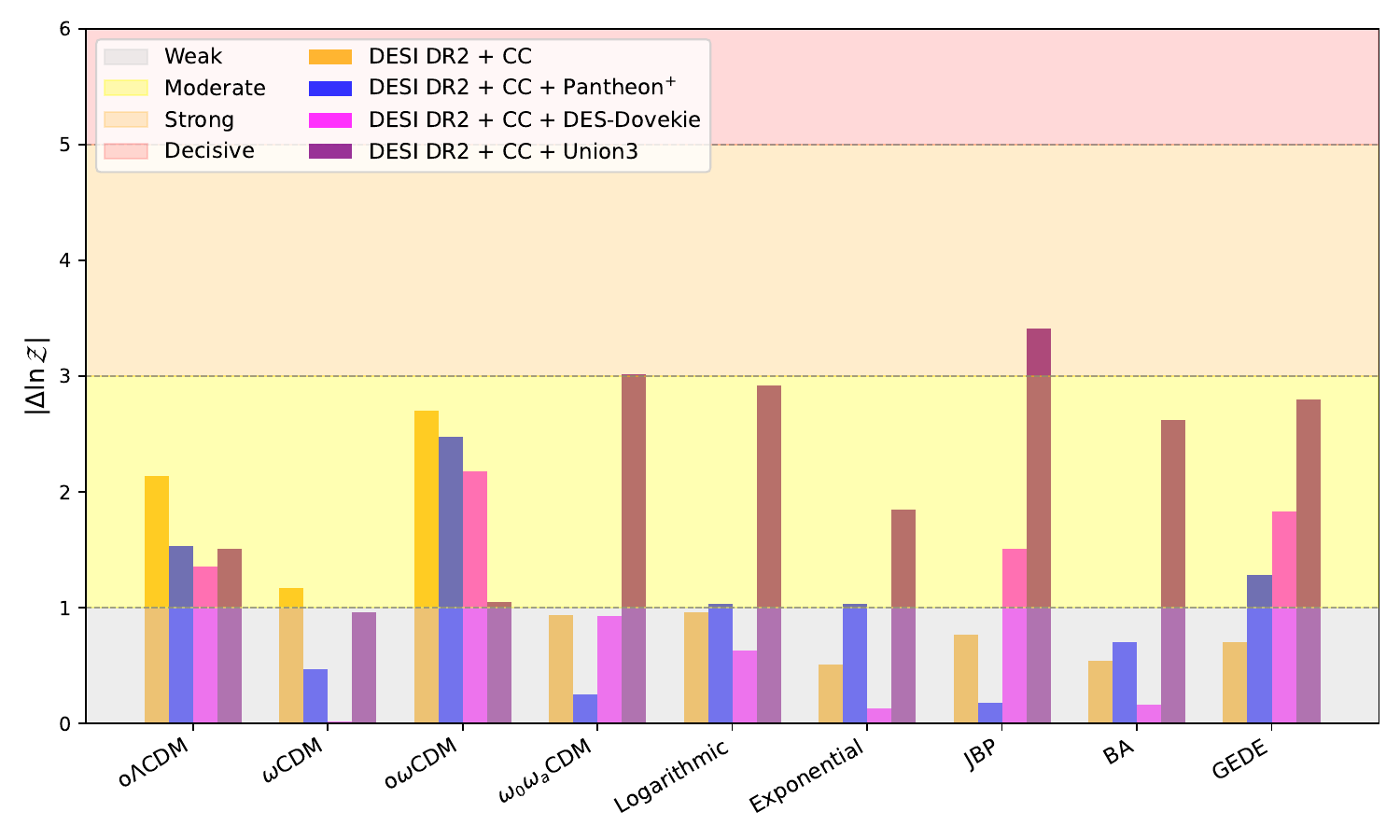}
\caption{This figure shows the difference in logarithmic Bayesian evidence of each dark energy model relative to the $\Lambda$CDM model using DESI DR2 with CC measurements and different SNe~Ia catalogs (Pantheon$^{+}$, DES-Dovekie, and Union3).}\label{fig_4}
\end{figure*}

\begin{figure*}
\centering
\includegraphics[scale=0.29]{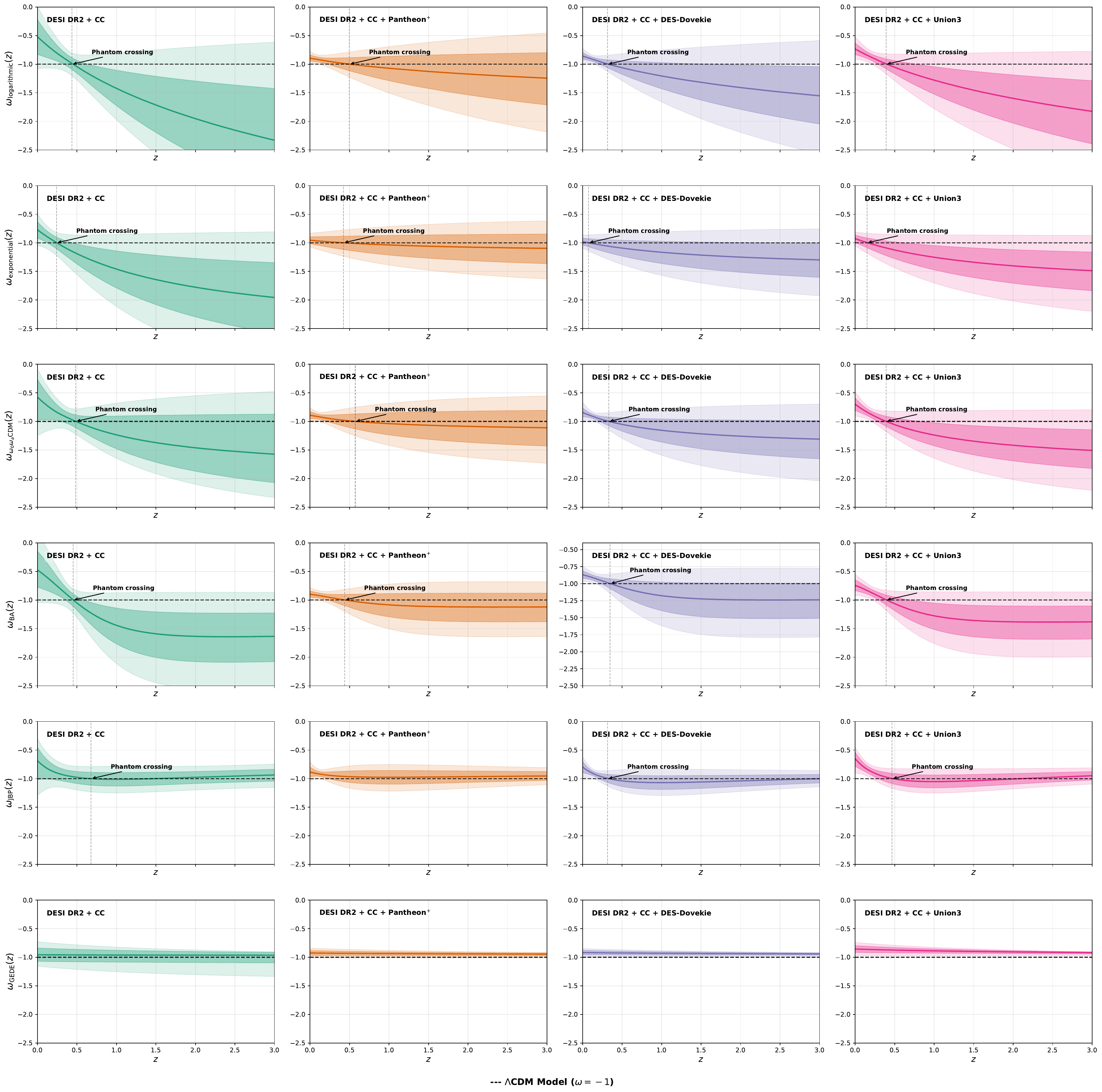}
\caption{This figure shows the difference in logarithmic Bayesian evidence of each dark energy model relative to the $\Lambda$CDM model using DESI DR2 with CC measurements and different SNe~Ia catalogs (Pantheon$^{+}$, DES-Dovekie, and Union3).}\label{fig_5}
\end{figure*}

\begin{figure*}
\centering
\includegraphics[scale=0.29]{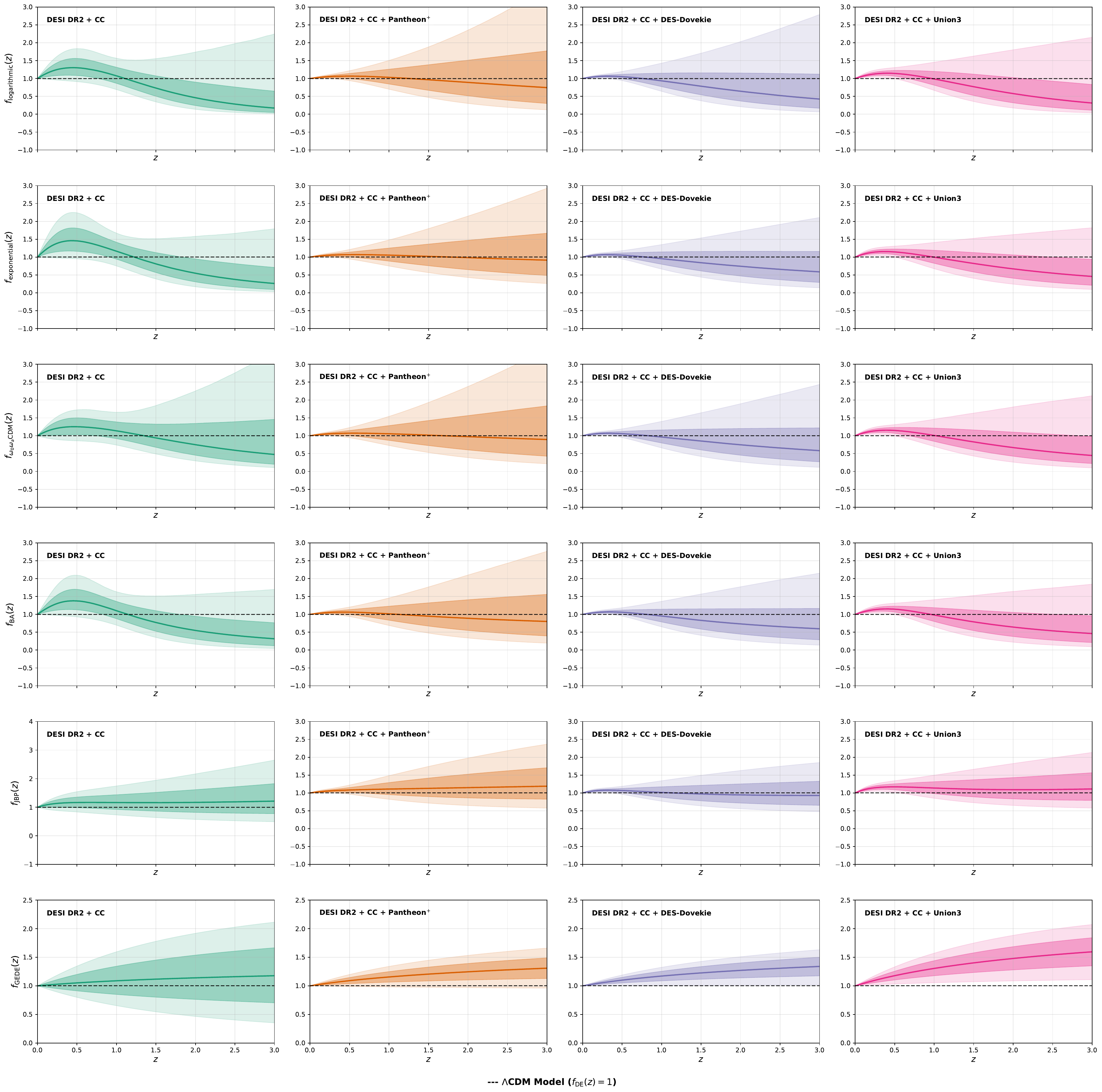}
\caption{This figure shows the evolution of $\omega(z)$ as a function of redshift, using DESI DR2 with CC measurements and different SNe~Ia catalogs (Pantheon$^{+}$, DES-Dovekie, and Union3).}\label{fig_6}
\end{figure*}

\section{Results}\label{sec_4}
In cosmology, most anomalies such as the Hubble tension, the $S_8$ discrepancy, the $M_B$ calibration offset, and the CMB lensing anomaly typically show deviations in the range of $2$–$4\sigma$. While not definitive, such levels are considered statistically significant and often prompt further investigation. Here, we carry out such an analysis by exploring a range of cosmological models and their compatibility with current observational datasets. To quantitatively assess departures from the $\Lambda$CDM reference model, we introduce a dimensionless tension estimator, $T \equiv \frac{\big| x_{\text{model}} - x_{\Lambda \mathrm{CDM}} \big|}
{\sqrt{\sigma_{\text{model}}^2 + \sigma_{\Lambda \mathrm{CDM}}^2}},$ where $x_{\text{model}}$ and $x_{\Lambda \mathrm{CDM}}$ denote the inferred values of a given cosmological parameter (such as $h$, $\Omega_m$, $\omega_0$, etc.) in the considered model and in $\Lambda$CDM, respectively, and $\sigma_{\text{model}}$ and $\sigma_{\Lambda \mathrm{CDM}}$ represent their corresponding uncertainties. The magnitude of $T$ is interpreted as follows: $T < 1\sigma$ indicates consistency with $\Lambda$CDM; $1\sigma \le T < 2\sigma$ corresponds to a weak and statistically insignificant deviation; $2\sigma \le T < 3\sigma$ reflects mild tension; $3\sigma \le T < 5\sigma$ indicates moderate tension; and $T \ge 5\sigma$ signifies a strong deviation.

Table~\ref{tab_3} shows the numerical values of the corresponding parameters for the $\Lambda$CDM, o$\Lambda$CDM, $\omega$CDM, $\omega_0\omega_a$CDM, Logarithmic, Exponential, JBP, BA and GEDE models using MCMC analysis. Fig.~\ref{fig_1} shows the corner plots of each cosmological model using combinations of DESI DR2, different SNe Ia catalogs and CC measurements. The diagonal panels show the 1D marginalized posterior distributions for each parameter, while the off-diagonal panels show the 2D marginalized confidence contours at 68\% and 95\% confidence levels.

First, we also obtain numerical values obtained from the CC measurements alone. For the $\Lambda$CDM model, we find $h = 0.662^{+0.053}_{-0.048}$ (hereafter referred to as the $\Lambda$CDM value), consistent with \cite{moresco2012improved,moresco2015raising,moresco20166}. Among its simple extensions, o$\Lambda$CDM predicts $h = 0.674 \pm 0.053$, corresponding to a $0.16\sigma$ deviation from the $\Lambda$CDM value, while GEDE predicts $h = 0.678 \pm 0.053$, deviating by $0.21\sigma$. $\omega$CDM yields $h = 0.689^{+0.087}_{-0.097}$, showing a slight tension of $0.24\sigma$ relative to the value of $\Lambda$CDM. In contrast, the o$\omega$CDM model predicts $h = 0.716^{+0.077}_{-0.089}$, showing about a $0.52\sigma$ deviation from the $\Lambda$CDM value. Similarly, the $\omega_0 \omega_a$CDM model yields $h = 0.727 \pm 0.082$, corresponding to a $0.67\sigma$ deviation, while the JBP model gives $h = 0.728^{+0.084}_{-0.072}$, with a deviation $0.74\sigma$. The BA parameterization predicts $h = 0.733 \pm 0.079$, marking the largest shift among these models at $0.75\sigma$. Among other parameterizations, the Logarithmic model estimates $h = 0.731 \pm 0.081$ ($0.71\sigma$ deviation), and the Exponential model provides the highest shift, $h = 0.736 \pm 0.079$, with a deviation $0.78\sigma$ relative to the $\Lambda$CDM value, although tensions remain below $1\sigma$, indicating consistency with $\Lambda$CDM.

We further analyzed the deviations of the Hubble parameter $h$ and $\Omega_{m0}$ for each dark energy model relative to the $\Lambda$CDM model, incorporating the CC dataset together with DESI DR2 and different SNe~Ia catalogs. The combined CC + DESI DR2 dataset shows good agreement with the $\Lambda$CDM model across most dark energy parameterizations. In particular, the o$\Lambda$CDM, $\omega$CDM, o$\omega$CDM, JBP, and GEDE models show deviations below the $1\sigma$ level in $h$ (with $T_h = 0.64\sigma$, $0.48\sigma$, $0.44\sigma$, $0.91\sigma$, and $0.40\sigma$, respectively), indicating that the corresponding values of the Hubble parameter $h$ remain fully consistent with the $\Lambda$CDM prediction. In contrast, the $\omega_0\omega_a$CDM, Logarithmic, Exponential, and BA parameterizations show weak and statistically insignificant tension in $h$, with deviations of $1.19\sigma$, $1.38\sigma$, $1.83\sigma$, and $1.57\sigma$, respectively.

For the matter density parameter $\Omega_{m0}$, a very similar behavior is observed. The o$\Lambda$CDM, $\omega$CDM, o$\omega$CDM, JBP, and GEDE models show deviations below the $1\sigma$ level (with $T_{\Omega_m} = 0.28\sigma$, $0.08\sigma$, $0.06\sigma$, $0.86\sigma$, and $0.08\sigma$, respectively), indicating that the inferred values of $\Omega_{m0}$ remain fully consistent with the $\Lambda$CDM prediction. In contrast, the $\omega_0\omega_a$CDM, Logarithmic, Exponential, and BA parameterizations shows weak and statistically insignificant tension in $\Omega_{m0}$, with deviations of $1.11\sigma$, $1.68\sigma$, $1.67\sigma$, and $1.62\sigma$, respectively.

For the CC + DESI DR2 + Pantheon$^{+}$ dataset, the $\omega_0\omega_a$CDM, Logarithmic, Exponential, JBP, and BA models show deviations below the $1\sigma$ level in $h$ (with $T_h = 0.56\sigma$, $0.56\sigma$, $0.62\sigma$, $0.86\sigma$, and $0.50\sigma$, respectively), indicating consistency with the $\Lambda$CDM model. In contrast, the o$\Lambda$CDM, $\omega$CDM, o$\omega$CDM, and GEDE models shows weak and statistically insignificant tension in $h$, with deviations of $1.13\sigma$, $1.22\sigma$, $1.11\sigma$, and $1.27\sigma$, respectively. For the matter density parameter $\Omega_{m0}$, the o$\Lambda$CDM, $\omega$CDM, o$\omega$CDM, $\omega_0\omega_a$CDM, Logarithmic, Exponential, JBP, BA, and GEDE models remain consistent with $\Lambda$CDM, with all lying below the $1\sigma$ level (with $T_{\Omega_m} = 0.83\sigma$, $0.53\sigma$, $0.22\sigma$, $0.13\sigma$, $0.13\sigma$, $0.07\sigma$, $0.12\sigma$, $0.26\sigma$, and $0.53\sigma$, respectively).

For the CC + DESI DR2 + DES-Dovekie dataset, the o$\Lambda$CDM, $\omega_0\omega_a$CDM, Logarithmic, Exponential, JBP, and BA models show deviations below the $1\sigma$ level in $h$ (with $T_h = 0.90\sigma$, $0.32\sigma$, $0.28\sigma$, $0.30\sigma$, $0.51\sigma$, and $0.29\sigma$, respectively), indicating consistency with the $\Lambda$CDM model. In contrast, the $\omega$CDM, o$\omega$CDM, and GEDE models show weak and statistically insignificant tension in $h$, with deviations of $1.38\sigma$, $1.48\sigma$, and $1.33\sigma$, respectively. For the matter density parameter $\Omega_{m0}$, the o$\Lambda$CDM, $\omega$CDM, o$\omega$CDM, $\omega_0\omega_a$CDM, Logarithmic, Exponential, JBP, BA, and GEDE models remain consistent with $\Lambda$CDM, with all deviations lying below the $1\sigma$ level (with $T_{\Omega_m} = 0.79\sigma$, $0.61\sigma$, $0.58\sigma$, $0.68\sigma$, $0.84\sigma$, $0.79\sigma$, $0.50\sigma$, $0.72\sigma$, and $0.56\sigma$, respectively).

Finally, for the CC + DESI DR2 + Union3 dataset, the $\omega_0\omega_a$CDM, Logarithmic, Exponential, and BA models show deviations of $0.95\sigma$, $0.92\sigma$, $0.99\sigma$, and $0.91\sigma$ in $h$, respectively, indicating consistency with $\Lambda$CDM. In contrast, the o$\Lambda$CDM, $\omega$CDM, o$\omega$CDM, JBP, and GEDE models shows weak and statistically insignificant tension in $h$, with deviations of $1.06\sigma$, $1.78\sigma$, $1.41\sigma$, $1.42\sigma$, and $1.18\sigma$, respectively. For the matter density parameter $\Omega_{m0}$, the o$\Lambda$CDM, $\omega$CDM, o$\omega$CDM, and GEDE models remain consistent with $\Lambda$CDM, with deviations of $0.66\sigma$, $0.42\sigma$, $0.35\sigma$, and $0.27\sigma$, respectively. In contrast, the $\omega_0\omega_a$CDM, Logarithmic, Exponential, JBP, and BA parameterizations show weak and statistically insignificant tension in $\Omega_{m0}$, with deviations of $1.67\sigma$, $1.65\sigma$, $1.65\sigma$, $1.11\sigma$, and $1.59\sigma$, respectively.

Fig.~\ref{fig_2} shows the different parameter planes of various cosmological models. These planes provide insights into the geometry of the Universe and the nature of dark energy. Panels (a), (b), and (c) of Fig~\ref{fig_2} show the $\omega-\Omega_m$, $\omega-\Omega_k$ and $\Omega_m-\Omega_k$ planes for the o$\Lambda$CDM, $\omega$CDM and o$\omega$CDM models, respectively. It is important to note that, in the case of the o$\Lambda$CDM model, for all combinations of the datasets the inferred curvature parameter is greater than zero, indicating a preference for an open Universe, which is compatible with recent late-time observations supporting an open Universe \citep{wu2025measuring}. A similar behavior is observed in the o$\omega$CDM model for the CC + DESI DR2 and CC + DESI DR2 + Pantheon$^{+}$ combinations. While in the o$\omega$CDM model, when the DES-Dovekie and Union3 datasets are included, the corresponding values are consistent with a spatially flat Universe, in agreement with the WMAP ($-0.0179 < \Omega_k < 0.0081$, 95\% CL) \citep{hinshaw2009five}, BOOMERanG ($0.88 < \Omega_{M/R} + \Omega_\Lambda < 1.0081$, 95\% CL) \citep{de2000flat}, and Planck ($\Omega_{M/R} + \Omega_\Lambda = 1.00 \pm 0.026$, 68\% CL) \citep{aghanim2020planck}

We also observe that in the cases of the $\omega$CDM and o$\omega$CDM models, the predicted value of $\omega$ in each case, $\omega_0 > -1$, shows a deviation from $\omega = -1$. However, in the o$\omega$CDM model, when we consider the combination CC + DESI DR2, we obtain $\omega_0 \approx -1$, which is close to the $\Lambda$CDM prediction. We also observe that in both cases the inclusion of different SNe Ia catalogs (Pantheon$^{+}$, DES-Dovekie, and Union3) shows a larger deviation from $\omega = -1$ compared to the CC + DESI DR2 combination alone.

Panels (d)–(h) of Fig~\ref{fig_2} show the $\omega_0-\omega_a$ planes for the $\omega_0\omega_a$CDM, Logarithmic, Exponential, JBP, and BA models, respectively. These provide important insights into the nature of dark energy. Each model predicts the values in the $\omega_0 > -1$ and $\omega_a < 0$ quadrant for each combination of DESI DR2 datasets with CC and different SNe~Ia calibrations, showing the preference for dark energy characterized by $\omega_0 > -1$, $\omega_a < 0$, and $\omega_0 + \omega_a < -1$, corresponding to a Quintom-B type behavior \citep{cai2025quintom,ye2025hints}, where the dark energy equation-of-state parameter transitions from $\omega < -1$ in the past to $\omega > -1$ at the present epoch.

On the other hand, panel (i) of Fig.~\ref{fig_2} shows the $\Delta$–$\Omega_m$ plane for the GEDE model, where the parameter $\Delta$ describes the evolution slope of the dark energy density, and the parameter $z_t$ denotes the transition redshift at which the dark energy density becomes equal to the matter density . Indeed, $z_t$ is not a free parameter but is determined by the condition that the dark energy density equals the matter density \citep{li2020evidence}. The GEDE model reduces to the $\Lambda$CDM model for $\Delta = 0$ and to the Phenomenologically Emergent Dark Energy (PEDE) model \citep{li2019simple} for $\Delta = 1$. It is crucial to note that, in each case, the GEDE model predicts a negative value of $\Delta$, indicating an injection of dark energy at high redshifts. Indeed, these results deviate from those reported by \cite{lodha2025desi,lodha2025extended,sharma2025probing} and \cite{chaudhary2025lambdacdm,chaudhary2025evidence}, since they consider the CMB in their analysis. Thus, one can also see the effect of late-time measurements on the predicted value of $\Delta$.

Fig~\ref{fig_3} shows the preference for dynamical dark energy over the cosmological constant ($\Lambda$), quantified in terms of the statistical significance departure from $\omega_0 = -1$. It can be observed that for the DESI DR2 + CC dataset, the $\omega$CDM and o$\omega$CDM models show deviations below $1\sigma$, while the $\omega_0\omega_a$CDM, Logarithmic, Exponential, JBP, and BA models show deviations of about $1.13\sigma$, $1.66\sigma$, $1.64\sigma$, $1.06\sigma$, and $1.71\sigma$, respectively. Indeed, all these deviations remain below the $2\sigma$ level, indicating the weak preference of dynamical dark energy. For the DESI DR2 + CC + Pantheon$^{+}$ dataset, the preference for dynamical dark energy increases noticeably for most models. The $\omega$CDM and $\omega_0\omega_a$CDM models both show deviations of about $1.75\sigma$, while the o$\omega$CDM model shows a deviation of $1.05\sigma$. The Logarithmic, JBP, and BA models show deviations at the level of $1.93\sigma$, $1.29\sigma$, and $1.82\sigma$, respectively, whereas the Exponential model shows the weakest deviation, remaining below $1\sigma$ at $0.74\sigma$.

For the DESI DR2 + CC + DES-Dovekie dataset, the preference for dynamical dark energy increases for most models. The $\omega$CDM, $\omega_0\omega_a$CDM, Logarithmic, JBP, and BA models show deviations of about $2.13\sigma$, $2.10\sigma$, $2.42\sigma$, $2.05\sigma$, and $2.43\sigma$, respectively, while the o$\omega$CDM model shows the smaller deviation of $1.62\sigma$. In contrast, the Exponential model shows a weaker preference for dynamical dark energy, remaining below $1\sigma$ at about $0.38\sigma$. For the DESI DR2 + CC + Union3 dataset, the $\omega$CDM, $\omega_0\omega_a$CDM, Logarithmic, JBP, and BA models show deviations of about $2.29\sigma$, $2.73\sigma$, $2.70\sigma$, $2.72\sigma$, and $2.74\sigma$, respectively, while the o$\omega$CDM model shows a preference of $1.96\sigma$. Once again, the Exponential model shows the weakest deviation, remaining below $1\sigma$ at $0.98\sigma$.

Finally, for the GEDE model, the preference for dynamical dark energy can be quantified by estimating the deviation from $\Delta = 0$. For the DESI DR2 + CC dataset, it shows a deviation of $0.27\sigma$. With the inclusion of the Pantheon$^{+}$ data, the GEDE model shows a deviation of about $1.80\sigma$, and for the DESI DR2 + CC + DES-Dovekie dataset, it shows a deviation of about $2.22\sigma$. For the DESI DR2 + CC + Union3 dataset, the deviation further increases and reaches $2.71\sigma$. Indeed, none of the cases reaches the $3\sigma$ level of deviation in the preference for dynamical dark energy. Although the DESI DR2 data show hints of new physics, the $\Lambda$CDM model remains consistent with the observations and is not yet ruled out by the late-time measurements.

Fig~\ref{fig_4} shows the comparative analysis of each dark energy model relative to the $\Lambda$CDM model based on the logarithmic bayesian evidence, using the revised Jeffreys’ scale. For the CC + DESI DR2 dataset, the o$\Lambda$CDM, $\omega$CDM, and o$\omega$CDM models show moderate evidence, while all other models show weak evidence. For the CC + DESI DR2 + Pantheon$^{+}$ combination, the o$\Lambda$CDM, o$\omega$CDM, Logarithmic, Exponential, and GEDE models show moderate evidence, whereas the remaining models show weak evidence. For the CC + DESI DR2 + DES-Dovekie combination, the o$\Lambda$CDM, o$\omega$CDM, JBP, and GEDE models show moderate evidence, while all other models remain weak evidence. Finally, for the CC + DESI DR2 + Union3 combination, the o$\Lambda$CDM, o$\omega$CDM, Exponential, BA, and GEDE models show moderate evidence, while the $\omega$CDM model shows weak evidence, and the $\omega_0\omega_a$CDM and JBP models show strong evidence. The corresponding numerical values associated with the logarithmic bayesian evidence can be seen in the eighth column of Table~\ref{tab_3}.

A complementary perspective is provided by the best-fit goodness-of-fit improvements, quantified by $\Delta \chi^2_{\min}$. For CC + DESI DR2, all models show only marginal improvement with $\Delta\chi^2_{\min} \lesssim 1.5$, indicating no strong deviation from $\Lambda$CDM, while the inclusion of Pantheon$^{+}$ slightly strengthens the preference for dynamical models, with $\Delta\chi^2_{\min} \approx 1.7$-$1.9$. The DES-Dovekie combination yields more pronounced improvements, with several models reaching $\Delta\chi^2_{_{\min}} \approx 3$, including $\omega_0\omega_a$CDM, Logarithmic, Exponential, JBP, and BA. The strongest gains arise for the Union3 dataset, most notably for $\omega_0\omega_a$CDM ($\Delta\chi^2_{\min} = 4.88$), followed by Logarithmic, Exponential, BA, and JBP. These $\Delta\chi^2_{\min}$ improvements are consistent with the Bayesian evidence results and further support the conclusion that Union3 provides the most compelling indication of departures from $\Lambda$CDM.

Fig~\ref{fig_5} shows the redshift evolution of the equation of state parameter $\omega(z)$ for each dark energy parameterization, using different combinations of datasets (DESI DR2, CC, Pantheon$^{+}$, DES-Dovekie, and Union3). In each model, with some exceptions for each combination of dataset, it can be observed that $\omega(z)$ falls below $\omega = -1$ at redshifts $z \gtrsim 0.5$, indicating a phantom regime ($\omega < -1$). At lower redshifts, around $z \lesssim 0.5$, $\omega(z)$ rises back above $-1$, entering the quintessence-like regime ($\omega > -1$). This crossing, where $\omega(z)$ evolves from the phantom to the quintessence regime by crossing $\omega = -1$, is known as phantom crossing, and this behavior, where the models show phantom behavior in the past and quintessence behavior in the present, is characterized by the quintom-B type scenario \citep{cai2025quintom,ye2025hints}. The first exception is the $\omega_0\omega_a$CDM model with the DESI DR2 + CC + Pantheon$^{+}$ combination, where the phantom crossing occurs around $z \approx 0.6$, while for the JBP model with the DESI DR2 + CC combination it occurs around $z \approx 0.65$. There is another expectation that, for each data combination, the mean value of $\omega(z)$ lies above $\omega = -1$ for the GEDE model and for the JBP model when combined with DESI DR2 + CC + Pantheon$^{+}$.

Fig~\ref{fig_6} shows the redshift evolution of the energy density $f_{DE}(z)$ presented in the third column of Table~\ref{tab_2} for each dark energy parameterization. It can be observed that, for all models, $f_{DE}(z)$ converges to 1 at $z=0$, i.e., $f_{DE}(0)=1$, for each combination of datasets. Also, for the GEDE model for each data combination and the JBP model with the DESI DR2 + CC, DESI DR2 + CC + Pantheon$^{+}$, and DESI DR2 + CC + Union3 combinations, the mean value of $f_{DE}$ lies above $f_{DE}=1$, while all models cross $f_{DE}=1$ at different redshifts depending on the chosen datasets. In Fig.~\ref{fig_5} and Fig.~\ref{fig_6}, the solid lines represent the mean values, while the light shaded and dark shaded regions correspond to the 1$\sigma$ and 2$\sigma$ confidence intervals, respectively.

\section{Discussion and Conclusions}\label{sec_5}
In this work, we have carried out a comprehensive Bayesian MCMC analysis of several cosmological models including $\Lambda$CDM, o$\Lambda$CDM, $\omega$CDM, $\omega_0\omega_a$CDM, Logarithmic, Exponential, JBP, BA, and GEDE using the most recent BAO measurements from more than 14 million galaxies and quasars drawn from the DESI Data Release 2 in combination with CC and different Type Ia supernova (SNe~Ia) catalogs (Pantheon$^+$, DES-Dovekie, and Union3). We constrain the key cosmological parameters ($h$, $\Omega_m$, $\Omega_k$, $\omega_0$, $\omega_a$, and $\Delta$) to investigate the preference for dynamical dark energy over the cosmological constant $\Lambda$.

Our results show that, using CC measurements alone, the $\Lambda$CDM model yields a reference value $h = 0.662^{+0.053}_{-0.048}$, while all dark energy models predict slightly higher values of $h$, with deviations remaining below the $1\sigma$ level. This indicates that CC measurements alone do not provide statistically significant evidence for departures from $\Lambda$CDM. When CC data are combined with DESI DR2, the constraints on $h$ and $\Omega_{m0}$ remain consistent with $\Lambda$CDM across most dark energy parameterizations. The o$\Lambda$CDM, $\omega$CDM, o$\omega$CDM, JBP, and GEDE models show deviations below the $1\sigma$ level in both parameters, while the $\omega_0\omega_a$CDM, Logarithmic, Exponential, and BA models show comparatively larger but statistically insignificant deviations, not exceeding the $\sim 2\sigma$ level. The inclusion of Type Ia supernova data (Pantheon$^{+}$, DES-Dovekie, and Union3) further improves the constraints without qualitatively altering these conclusions. In all cases, $\Omega_{m0}$ remains fully consistent with $\Lambda$CDM at the $1\sigma$ level, while mild, dataset-dependent deviations in $h$ at the $\sim 1$-$2\sigma$ level are observed for some models, remaining statistically insignificant.

Our results show that the o$\Lambda$CDM model favors an open Universe for all combinations of the datasets. In contrast, for the o$\omega$CDM model, the CC + DESI DR2 and CC + DESI DR2 + Pantheon$^{+}$ combinations again favor an open Universe, while the CC + DESI DR2 + DES-Dovekie and CC + DESI DR2 + Union3 datasets favor a nearly flat Universe, supporting a spatially flat geometry consistent with WMAP, BOOMERanG, and Planck observations. Furthermore, the joint posterior analyses of the $(\omega_0, \omega_a)$ parameter planes reveal that most dynamical dark energy parameterizations favor $\omega_0 > -1$ and $\omega_a < 0$, indicating that such dynamical dark energy is characterized by a Quintom-B–type behavior.

Consequently, our analysis shows that using late-time observations, dark energy extensions beyond the cosmological constant exhibit deviations from $\Lambda$CDM below the $1\sigma$ level for the $\omega$CDM, o$\omega$CDM, and GEDE models when the DESI DR2 + CC dataset is used, while the remaining dark energy models show deviations that remain below the $2\sigma$ level. In contrast, when the Pantheon$^{+}$ SNe Ia dataset is included, most dynamical dark energy models exhibit deviations from $\Lambda$CDM at the $1\sigma$-$2\sigma$ level when combined with DESI DR2 + CC data, while the o$\omega$CDM and Exponential models remain below the $1\sigma$ level. With the inclusion of the DES-Dovekie and Union3 SNe Ia datasets, most dynamical dark energy models show deviations from $\Lambda$CDM at the $2\sigma$-$3\sigma$ level when combined with DESI DR2 + CC data, while the o$\omega$CDM model remains below the $2\sigma$ level and the Exponential model consistently shows deviations below $1\sigma$. Indeed, these deviations show that the preference for dynamically dark energy reaches the $2$-$3\sigma$ level; yet, it remains insufficient to conclusively rule out the $\Lambda$CDM model.

Bayesian evidence and goodness-of-fit analyses show that the level of support for dark energy extensions beyond $\Lambda$CDM depends strongly on the chosen dataset. While the CC + DESI DR2 combination provides at most moderate support for o$\Lambda$CDM and o$\omega$CDM, with only small improvements in the fit ($\Delta\chi^2_{\min} \lesssim 1.5$), the inclusion of the Pantheon$^{+}$ sample slightly increases this preference ($\Delta\chi^2_{\min} \simeq 1.7$-$1.9$). In contrast, the DES-SN5Y and Union3 datasets give the strongest support for dynamical dark energy models, such as $\omega_0\omega_a$CDM, JBP, BA, and GEDE, and show larger improvements in the fit, including the highest $\Delta\chi^2_{min}$ values.

The evolution of the equation of state parameter $\omega(z)$ shows that most dark energy parameterizations undergo a transition from a phantom regime ($\omega < -1$) at intermediate redshifts to a quintessence-like regime ($\omega > -1$) at low redshifts, consistent with a phantom-crossing, quintom-B type behavior, with the transition redshift depending on the model and dataset. Notable exceptions include the GEDE model and the JBP model when combined with DESI DR2 + CC + Pantheon$^{+}$, for which the mean value of $\omega(z)$ remains above $-1$. The corresponding evolution of the normalized dark energy density $f_{DE}(z)$ satisfies $f_{DE}(0)=1$ for all cases and exhibits dataset-dependent crossings of $f_{DE}=1$, supporting a time-varying dark energy component compatible with current observational constraints.

In conclusion, our analysis indicates that the standard $\Lambda$CDM cosmology remains a statistically robust description of the current Universe, although late-time observational datasets, particularly DESI DR2 and recent SNe~Ia samples, provide moderate evidence supporting the possibility of dynamical dark energy models. Future Stage IV surveys and next generation observatories are poised to significantly advance our understanding of the dark sector. DESI will deliver refined constraints from DR2 probes such as full-shape fitting, bispectrum, gravitational lensing, and peculiar velocities in 2025–2026, with DR3 results expected in 2027, which may shed light on the possible phantom crossing~\citep{ade2019simons,dawson2022snowmass2021} and the extent of any deviations from the $\Lambda$CDM model. The Nancy Grace Roman Space Telescope, launching in 2026, and the proposed DESI-II in the 2030s will probe the $z > 1$ Universe, opening a new frontier for dark energy studies~\citep{spergel2015wide,dawson2022snowmass2021}. If Stage IV surveys break $\Lambda$CDM model, developing new observational methods, cross-survey analyses, and searches for specific model signatures will be crucial, and a dedicated post $\Lambda$CDM dark energy task force could provide guidance for exploring the complex physics of the dark sector.

\section*{Acknowledgements}
SC acknowledges the Istituto Nazionale di Fisica Nucleare (INFN) Sez. di Napoli,  Iniziative Specifiche QGSKY and MoonLight-2  and the Istituto Nazionale di Alta Matematica (INdAM), gruppo GNFM, for the support. This paper is based upon work from COST Action CA21136 -- Addressing observational tensions in cosmology with systematics and fundamental physics (CosmoVerse), supported by COST (European Cooperation in Science and Technology). VKS gratefully acknowledges the facilities and institutional support provided by the Indian Institute of Astrophysics (IIA), India, during his tenure as a postdoctoral fellow.

\bibliographystyle{elsarticle-num}
\bibliography{mybib.bib}

\end{document}